\documentclass{cernrep}

\def\half{{1 \over 2}}
\def\({\left(}
\def\){\right)}
\def\[{\left[}
\def\]{\right]}
\def\k{{\bf k}}
\def\x{{\bf x}}

\def\ltap{\raisebox{-.55ex}{\rlap{$\sim$}} \raisebox{.4ex}{$<$}}
\def\gtap{\raisebox{-.55ex}{\rlap{$\sim$}} \raisebox{.4ex}{$>$}}

\def\lsim{\mathrel{\ltap}}
\def\agt{\mathrel{\gtap}}

\newcommand{\e}{\mathop{\rm e}\nolimits}

\begin{document}
\title{Cosmology and Dark Matter}
\author{I. Tkachev}
\institute{ Institute for Nuclear Research of the Russian Academy of Sciences, Moscow, Russia}

\maketitle

\begin{abstract} 
This lecture course covers cosmology from the particle physicist perspective.
Therefore, the  emphasis will be on the  evidence for the new physics  in cosmological and astrophysical data together with minimal theoretical frameworks needed to understand and appreciate the evidence. I review the case for non-baryonic dark matter and describe popular models which incorporate it. In parallel, the story of dark energy will be  developed, which includes accelerated expansion of the Universe today, the Universe origin in the Big Bang, and support for the Inflationary theory in CMBR data.
\end{abstract}


\section{Introduction}

The deeper we dig into microphysics, the deeper connections between smallest and largest scales are reviled. The world which surrounds us on Earth in everyday life can be understood in frameworks of electrodynamics with atoms and molecules in hands. But  we need nuclear physics to describe the  Sun and other stars. Moving to even bigger scales, explaining  galaxies and the Universe as a whole, we need good understanding of particle physics.

This relation works the other way around as well.  In particular, cosmology  tells us that the Standard Model of particle physics is incomplete. Namely, that the galaxies and galaxy clusters are made mainly of the Dark Matter, which overweights usual baryonic matter, and for which there is no room in the Standard Model of particle physics. It tells also that at the scale of the Universe the Dark Energy overrules, which can be simple or complicated substance, but which is not matter. We conclude that there ought to be a new physics and new particles outside of the Standard Model frameworks. At present we learn their properties from cosmology only. 

The plan of this lectures is as follows. In Section~\ref{sec:CosmologyBasics},
 I review the basics of cosmology: Friedman equations, Hubble expansion and cosmography.  Evidence for the dark energy is also presented in this section.
In Section~\ref{sec:HBB} the Hot Big Bang theory is outlined: relevant thermodynamal facts and relations are presented and {Cosmic Microwave Background Radiation} (CMBR) is introduced.
 ln Section~\ref{sec:CMBR_tool} the study of CMBR anisotropies is described as a tool of precision cosmology. In Section~\ref{sec:inflation} basics of inflationary theory are given, while some important technical details of this theory are moved into Appendices. In Section~\ref{sec:DM} the  evidence for dark matter is described together with dark matter models and dark matter searches results.

\section{Basics of Cosmology}
\label{sec:CosmologyBasics}

There are many excellent recent books and reviews on cosmology in the market, which readers may consult for missing details. I would especially recommend the balanced, contemporary and comprehensive book by Gorbunov and Rubakov\cite{RubakovGorbunov}.

\subsection{Expansion of the Universe}

Cosmological dynamics is provided by General Relativity - the Einstein field equations
\begin{equation}
R_{\mu \nu} -\half g_{\mu \nu} R = 8\pi G \; T_{\mu \nu}\; ,
\label{Einstein-equations}
\end{equation} 
where $T_{\mu \nu}$ is a stress energy tensor describing the distribution of
mass in space, $G$ is Newton's gravitational constant, and the curvature
$R_{\mu \nu}$ is certain function of the metric $g_{\mu\nu}$ and its first and second
derivatives. Immediate consequence of Einstein equations is energy momentum conservation
\begin{equation}
T^{~\nu}_{\mu\,\,\,\,;\nu}=0 \; .  
\label{relativ-energy-conservation}
\end{equation}
These equations take simple form in important physical situations with special symmetries.
At large scales the Universe as a whole is homogeneous and isotropic and these symmetries form basis for 
the construction of cosmological models.  The most general space-time metric describing such universe
is the Robertson-Walker metric
\begin{equation}
ds^{2} = dt^{2} - a^{2}(t)\; {\bf dl}^{2}\; ,
\label{FRW-metric}
\end{equation} 
where $a(t)$ is the dimensionless scale factor by which all distances vary as
a function of cosmic time $t$.  The scale factor contains all the dynamics of
the Universe as a whole, while the vector product ${\bf dl}^{2}$ describes the geometry
of the space,
$$
{\bf dl}^{2} = \frac{dr^{2}}{1-k\,r^{2}} + r^{2}\,(d\theta^{2} +
\sin^{2}\theta \, d\phi^{2})\; ,
$$
which can be either Euclidian, or positively or negatively curved.
For the spatial 3-dimensional curvature we find, explicitly
\begin{equation}
{}^{(3)}\!R = {6k\over a^2(t)}\; \; \; 
\left\{\begin{array}{ll}
k=-1&\hspace{0.5cm}{\rm   Open} \\
k=0&\hspace{0.5cm}{\rm Flat}\\
  k=+1&\hspace{0.5cm}{\rm   Closed} 
\end{array}\right.
\label{Spatial-curvature} 
\end{equation} 
E.g., the space with $k=+1$ can be thought of as a 3-dimensional sphere
with a curvature being inversely proportional to the square of its radius.  In
this Section we will model the matter content of the Universe as a perfect
fluid with energy density $\rho$ and pressure $p$, for which the stress-energy
tensor in the rest frame of the fluid is
\begin{eqnarray}
T_{\mu}^{~\nu}=\left( \begin{array}{cccc}
{\rho} & 0 & 0 & 0\\
0 & {-p} & 0 & 0 \\
0 & 0 & {-p} & 0 \\
0 & 0 & 0 &{-p}
\end{array} \right) 
\label{Tmunu-ideal}
\end{eqnarray} 
With these assumptions the Einstein equations simplify to the Friedmann
equations, which form the dynamical basis of cosmology 
\begin{eqnarray} \label{eq:fridman1}
\frac{\dot{a}^{2}}{a^{2}} &=& \frac{8\pi G}{3}\, \rho ~-~
\frac{k}{a^{2}} \; , \\ 
\frac{\ddot{a}}{a}~ &=& -~ \frac{4\pi G}{3}\; (\rho + 3\, p) \; .
 \label{eq:fridman2}
\end{eqnarray}

As Alexander Friedmann have shown in 1922, a universe described by such equations cannot be static, it inevitably expands or collapses. Solution of Eqs.~(\ref{eq:fridman1}) and (\ref{eq:fridman2}) can be found in the following way, which also highlights the physics behind these equations.  Differentiating the first  Friedmann equation and combining result with the second one we obtain  
\begin{equation} 
\frac{d{\rho}}{dt} ~+~ 3\,\frac{\dot{a}}{a}\, ({\rho}+{p})=0, 
\label{FirstLawTherm}
\end{equation}
which is nothing but the energy-momentum conseravtion  $T^{\mu\nu}_{\,\,\,\,\,\,\,\,\,;\nu}=0$ written for the homogeneous isotropic medium. On the other hand the result Eq.~(\ref{FirstLawTherm}) also corresponds to the First Law of thermodynamics
\begin{equation}
dE ~+~ p\, dV ~=~ T\, dS,
\label{FirstLawTherm2}
\end{equation}
with $dS=0$, where $S$ is entropy, $ E = \rho V$,  and $ V \propto a^{3}$. This is expected since in derivation of Friedmann equations the energy-momentum tensor of an ideal fluid was assumed. It turns out that this is valid approximation most of the time\footnote{It is violated though during special moments, in particular at the initial matter creation after inflation, see below, or at strongly first order phase transitions, if those existed in the Universe past.} and therefore the expansion of the Universe is adiabatic, $S = \rm const$.

Let $s$ be entropy  density,  $S = s V$. We know from thermodynamic (and I'll derive this in Section~\ref{sec:ThermDynUniv}) that in thermal equilibrium the entropy  density is given by
\begin{equation}
s = {2\pi^2\over45}\, g_*\; T^3,
\label{eq:entropy}
\end{equation}
where $g_*$ is the number of effective relativistic degrees of freedom
\begin{equation}
\footnotesize  g_* = {\footnotesize \sum_{i={\rm bosons}} g_i +
{7\over8}\sum_{j={\rm fermions}} g_j},
\label{eq:g*}
\end{equation}
and $g_i$ is the number of spin states of a given particle, e.g. for photons, electrons and positrons $g_\gamma = g_{e^-} = g_{e^+} = 2$.
In this expression only particles with $  m \ll T$ are counted, i.e. $g_*$
is a function of temperature. It is displayed in Fig.~\ref{fig:gT} by the solid line for the Standard Model and by the dashed line for the minimal supersymmetric extension of the SM. Therefore, $S = \rm const$ is  equivalent to 
\begin{equation}  g_*^{1/3}\,  T \, a= {\rm const}.
\label{eq:aT}
\end{equation}
This is very useful relation. In particular, it gives   $T \propto a^{-1}$  (neglecting the change in the number of  degrees of freedom), and we arrive to the concept of the Hot Universe right away. Indeed, currently the Universe is expanding, in the past it was smaller and therefore hotter. In the era of  precision cosmology the change in $g_*(T)$  should be counted too, if we go beyond simple estimates. We do not know all particle content of complete theory describing Nature, but LHC will fix the actual shape of $g_*(T)$ in the region of highest $T$ of Fig.~\ref{fig:gT} and, in fact,  a sharp MSSM like  rise in the number of degrees of freedom is ruled out already.

Only relativistic particles contribute to entropy, but everything existing contributes to  the energy density, $\rho$. Even vacuum. The equation of state, $w$, of a substance contributing to $\rho$ is defined as
$ w \equiv {p}/{\rho} $. If $w = \rm const$,  the energy-momentum conservation, Eq. (\ref{FirstLawTherm}),
gives $\rho = a^{-3(1+w)}\; \rho_0 $. Using this result, we find from the first Friedmann equation, Eq.~(\ref{eq:fridman1}), the scale factor as a function of time, $a = \( {t}/{t_0} \) ^{{2}/{3(1+w)}}$. 

During the first half of of the last century cosmologists were assuming that the Universe is filled with a ``dust'', $p=0$, while the ``dust'' particles were represented by galaxies made of usual matter. Nowadays we know that the Universe is multicomponent.
Its energy density  was dominated in turn by radiation ($p = \rho/3$), by dark matter  ($p = 0$, as for the "dust" of galaxies) and finally by dark energy  (equation of state consistent with $p = -\rho$). 
In the Table \ref{tab:EqSate} we list: substances known to contribute into energy balance in the Universe, their defining equations of state, the corresponding scaling of energy density with expansion, and corresponding solution for $a(t)$ if $k=0$.
\begin{table}[h]
\begin{center}
\caption{Substances contributing into the energy balance in the Universe}
\label{tab:EqSate}
\begin{tabular}{|l|l|c|c|}
\hline
Substance & Equation of state~~ & $\rho(a)$ & $a(t)$ \\
\hline
Radiation~~~~~~~   & $ w = {1}/{3}$ ~~~            & $\rho = a^{-4}\; \rho_0 $ & $a = (t/t_0)^{1/2} $\\
Matter   & $ w = 0$             & $\rho = a^{-3}\; \rho_0 $ & $a = (t/t_0)^{2/3}$ \\
Vacuum   & $ w = -1$             & ~~$\rho = \rm const~~~~~ $ & ~~$a = \exp{(H_0 t)}$\\
\hline
\end{tabular}
\end{center}
\end{table}

To parameterize the Fiedmann equations and their solution $a(t)$, cosmologists introduce {\it cosmological parameters}. One of such parameters we have already encountered, this is $k$ entering Frideman equations explicitly. Despite paramtrizing spatial geometry of the Universe it was used to predict the Universe future. Namely, it immediately follows from  Eq.~(\ref{eq:fridman1}) that for $ k = - 1$ or $ k = 0$ the  Universe will expand forever. For $k = + 1$ the Universe should recollapce at the point when r.h.s. of  Eq.~(\ref{eq:fridman1}) turns to zero. This should happen in radiation or matter dominated Universe, but never happens in the universe dominated by the dark energy with $\rho_0 > 0$. Since nowadays we know that our Universe is dominated by such dark energy, we already know that it will expand forever. 

Inherently related to the parameter $k$ is   {\it critical density}. This is the density at which the Universe is spatially flat,  $k=0$
\begin{equation}
\rho_c \equiv \frac{3}{8\pi G}\; \(\frac{\dot{a}}{a}\)^{2}.
\label{eq:CritRho}
\end{equation}
Critical density  can be expressed using  a second, directly observable and very important cosmological parameter, the Hubble constant, 
$H \equiv {\dot{a}}/{a} $.
To quantify relative contribution, $\rho_i$, of each of the components in the total energy budget of the Universe, $\rho$, the following notations are introduced, $\Omega_i \equiv \rho_i /\rho_c$ and $\sum_i \rho_i = \rho$.
 The current knowledge of the numerical values of some of these parameters at $t=t_0$ is summarized in the Table~\ref{tbl:cosmoparameters} below.

\begin{table}[h]
\begin{center}
\caption{Cosmological parameters}
\begin{tabular}{|l|l|l|}\hline
Symbol \& Definition & Description & Present value, from Ref.~\cite{Ade:2015xua}\\
\hline
$t$                & {Age of the Universe} & $ t_0 = 13.81 \pm 0.03 $ ~Gyr\\ 
$H = \dot{a}/a$    & {Hubble parameter} & $ H_0 = 67.27 \pm 0.66 \rm {~km~ s^{-1}~ Mpc^{-1}}$\\
$\Omega = \rho/\rho_c$  & Spatial curvature &  $  1-\Omega  = 0.000 \pm 0.005$  \\
$\Omega^{}_\gamma = \rho^{}_\gamma/\rho_c$ & Fraction of photons &
$\Omega^{}_\gamma = 2.48\cdot 10^{-5}\; h^{-2}$ \\
$\Omega_b = \rho_b/\rho_c$ &  Baryonic fraction & $\Omega_b\, h^2= 0.02225 \pm 0.00016$ \\
$\Omega_{\rm m} = \rho_{\rm m}/\rho_c$ &  Matter fraction & $\Omega_{\rm m} = 0.316 \pm 0.009$ \\
$\Omega_\Lambda = \rho_\Lambda/\rho_c$ &  Dark Energy fraction &
$\Omega_\Lambda = 0.684 \pm 0.009$ \\
\hline
\end{tabular}
\label{tbl:cosmoparameters}
\end{center}
\end{table}

Accuracy of numerical values  presented in this table should not be over-appreciated since those were derived with some model assumptions, e.g. that the dark matter is cold and the equation of state for dark energy is $w=-1$. Relaxing such assumptions changes presented values somewhat.

\subsection{The Hubble law}

   The velocity with which  distance $r = a(t)r_0$ between two arbitrary galaxies increases in expanding Universe  follows trivially  from the definition of the Hubble parameter, $ H \equiv \dot{a}/a$, namely $  ~v =  \dot{r} = H\, r$.   This relation, known as the  Hubble law, makes the  basis for direct observational determination of the Hubble parameter and was discoverd by Edwin Hubble in 1929. 
It is also used to set  units for measuring $H$.  For convenience, in many cosmological relations,  dimensionless ``small $h$'' is  introduced as {\small $H = 100\, h\, \rm ~km~s^{-1} Mpc^{-1}$}, then $h \sim O(1)$.  Latest value of the Hubble constant obtained from direct mapping of recession velocities versus distance corresponds to $H_0 = (73.00 \pm 1.75) {\rm ~km~ s}^{-1} {\rm ~Mpc}^{-1}$ \cite{Riess:2016jrr}. Note that this value is $3.4\sigma$ higher than the value presented in Table~\ref{tbl:cosmoparameters} which was derived indirectly from other cosmological data. This may suggest some unaccounted systematic uncertainties or may indicate new physics~\cite{Berezhiani:2015yta}.

Looking at  cosmologically distant objects we see just the light they emit. One can  wander  then,
how observables are derived from this limited information?
E.g. how distance and velocity can be measured separately to determine $H$? The unswear is simple. Velocity is measured by the frequency shift of known signal, similarly to what police is doing  when checking for speeding cars using Doppler radars. Distance can be derived measuring dimming of objects with calibrated intrinsic luminosity: objects which are further away are less bright.  Now we shall explain this in more details.

\subsubsection{Redshift}

Photon motion in any metric is described by basic equation, $ds^2 = 0$.  In the Robertson-Walker metric this becomes $ ds^{2} = dt^{2} - a^{2} d\chi^{2} = 0$, where $\chi$ is {\it comoving} distance along particle trajectory.
Define  conformal time ~$ \eta$ ~as ~
\begin{equation}
d\eta = dt/ a. 
\label{eq:time-conformal}
\end{equation}
Then 
$ds^{2} = a^{2}(d\eta^{2} - d\chi^{2}) = 0$. Remarkably, solution for a photon world line in conformal coordinates is the same as in Minkowsikan space-time $\chi = \pm \eta \;+\; {\rm const}$. Therefore, the conformal time lapse between two events of light emission at one point will be the same as for their detection at another point, regardless of distance traveled, $d \eta|_{\rm emission} = d \eta|_{\rm detection}$. Therefore, the proper time lapses for emissions and detections are related  as
$$\large \frac{dt}{a}|_{\rm detection} ~=~ \frac{dt}{a}|_{\rm emission}.$$
Let $dt$ corresponds to the period of some  monochromatic signal, then for its frequency we will have
$\omega \, a_d ~=~ \omega_0 \, a_e$. 
As a  result, we can say that the wavelength of a signal stretches together with the expansion of the Universe. An this interpretation is often used. However, it is incorrect. For example, the space does not stretches inside galaxies, but $\omega \, a_d ~=~ \omega_0 \, a_e$ will be always true regardless of how many times the light signal passed through galaxies between detection and emission. The signal frequency changes between the point of emission, $\omega_0$, and in the point of detection, $\omega$, because clocks run differently in those points. In a similar way the gravitational redshift or blueshift can be also derived and interpreted. At largest cosmological scales we always have redshift since Universe is expanding.

In measurements, the redshift is quantified as $z ~\equiv~ {(\omega_0 - \omega)}/{\omega} $ and measured as a frequency shift of emission or absorption lines of various chemical elements. With this definition we get 
\begin{equation}
1+z = \frac{a_d}{a_e}. 
\label{eq:a_z}
\end{equation}
In other words, redshift can be used also to label cosmological epoch.
For convenience, we can always normalise the scale factor today as $a_d = 1$.
Recall relation $a \propto T^{-1}$ (or, more precisely, Eq.~\ref{eq:aT}). This gives temperature of the Universe as a function of redshift, $T(z) = T_0 (1+z)$. Also, using our previous result $\rho = a^{-3(1+w)}\; \rho_0 $ and Friedmann equation (\ref{eq:fridman1}) we can rewrite Hubble parameter as a function of $z$ and cosmological parameters $H_0$ and $\Omega_i$ in the following often used form
\begin{equation}
H^{2} (z) = H_0^{2}\, \sum_i  \Omega_i\, (1+z)^{3(1+w_i)} .
\label{eq:H_z}
\end{equation}
Differentiating Eq.~(\ref{eq:a_z}) we find $dz = - H d\chi $ which  is a local form of the Hubble law, partially expressed through observables already. Instead of $v$ we have directly measurable redshift z. This is understandable since  in the non-relativistic limit cosmological redshift reduces to the Doppler effect - at the end, galaxies are receding from us with the Universe expansion.  Finally, integral of the local form of the Hubble law, $dz = - H d\chi $, gives
\begin{equation}
\chi (z) =   \int_0^{z} \frac{dz'}{H(z')} \, . 
\label{eq:chi_z}
\end{equation}
Now we have to find the way to measure distances to remote objects to complete construction of the Hubble law generalisation, which would be expressed through observables only and which would be valid to any redshift. 

\subsubsection{Luminosity distance}

Consider two objects with identical luminosities (Standard Candles) placed at different distances from us.
The radiation flux, $F$, scales with distance as $F^{-1/2}$. Therefore, measuring fluxes from a standard candles, we can determine the ratio of distances to them. Moreover, the distance to an object with known luminosity can be defined as Luminosity Distance, $ D_L$, and measured via measuring flux from it
 $$ D_L^{2} ~\equiv~ \frac{L}{4\pi F} .$$

Consider now this idea in cosmology. Again, let us write metric in conformal time, but now we should keep track of changing area with distance at fixed solid angle,  $ds^{2} = a^{2}\;(d\eta^{2} - d\chi^{2} - \chi^{2}\; d\Omega)$.
Surface area at the point of detection  is ~$4\pi\, \chi^{2}$ (we can always normalise scale factor today as $a = 1$). Further,  energy and arrival rates of registered photons are redshifted. This reduces the flux by
$(1+z)^{2}$, where $z$ is redshift of emitter. We get for observed bolometric flux 
$$F ~=~ \frac{L}{4\pi\, \chi^{2}(1+z)^{2}},$$
and luminosity distance is
$$ D_L ~\equiv~ \sqrt{\frac{L}{4\pi F}} ~=~ (1+z)\; \chi  .$$
Therefore, measuring flux, we can determine comoving distance to a standard candle. We may not know the value of intrinsic luminosity $L$, but this is not necessary. It is important only that $L$ should not vary from an object to object. Then we can compare ratio of fluxes at different redshifts and from there derive cosmology. For historical reasons 
astronomers measure flux in magnitudes,  which are defined as $\mu \propto 5\log_{10} F$.  Ratio of fluxes will be difference in magnitudes. Now, we want to see how different are magnitudes of a standard candles at a given redshift in different cosmologies, say in cosmology which predicts $\chi(z)$ and in a ``base'' cosmology which predicts $\chi(z)_{\rm base}$. We find
\begin{equation}
\Delta\mu=\mu-\mu_{\rm base} = 5\log_{10} \[\frac{\chi(z)}{\chi(z)_{\rm base}}\]
\label{eq:Dmu}
\end{equation}
Here $\chi (z)$ and $\chi(z)_{\rm base}$ are given by Eq.~(\ref{eq:chi_z}) with its own sets of cosmological parameters each. 

\subsubsection{Dark Energy}
\label{sec:SNLumDist}

\begin{figure}
\begin{center}
\includegraphics[width=0.54\textwidth]{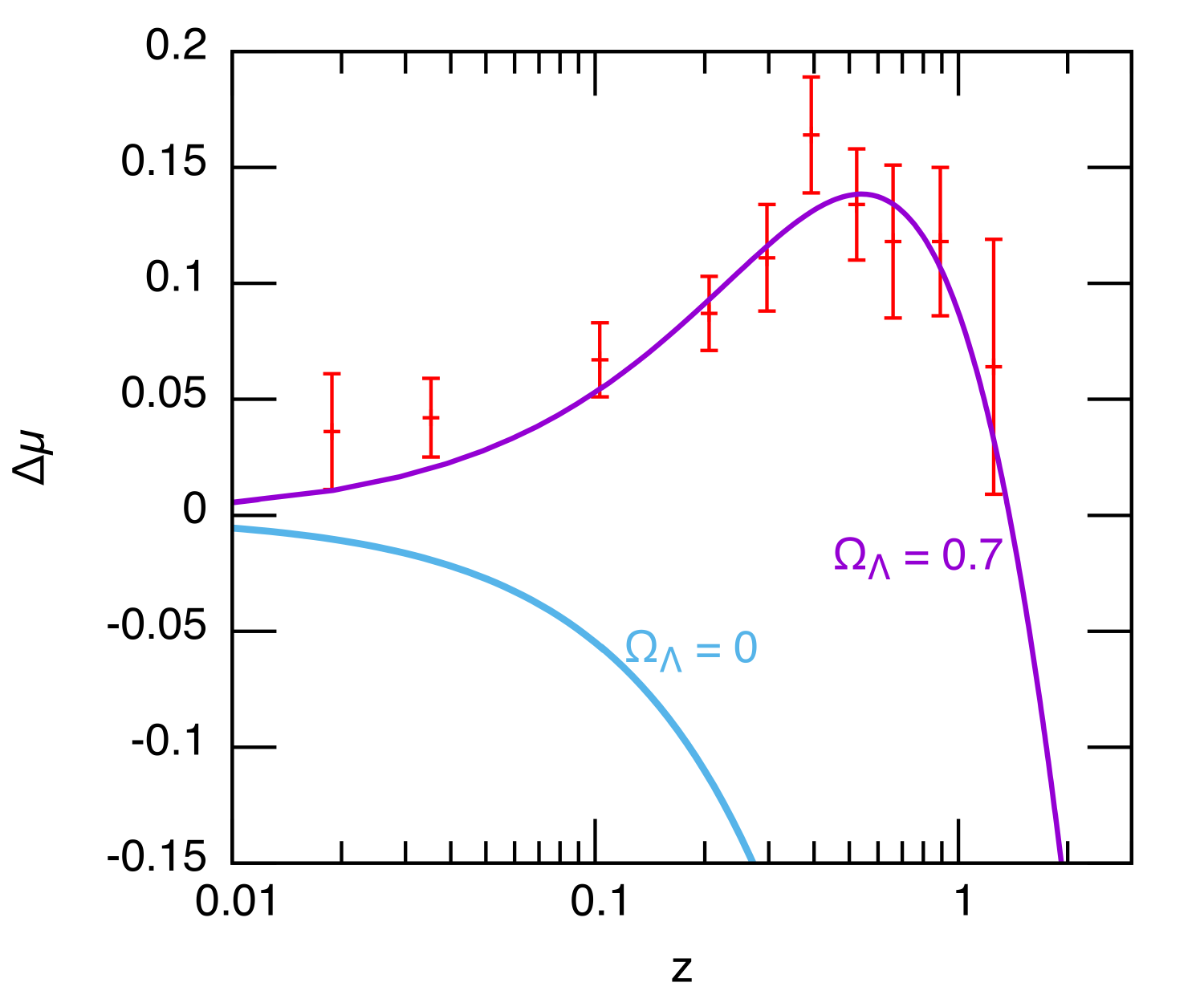}~~~~
\includegraphics[width=0.44\textwidth]{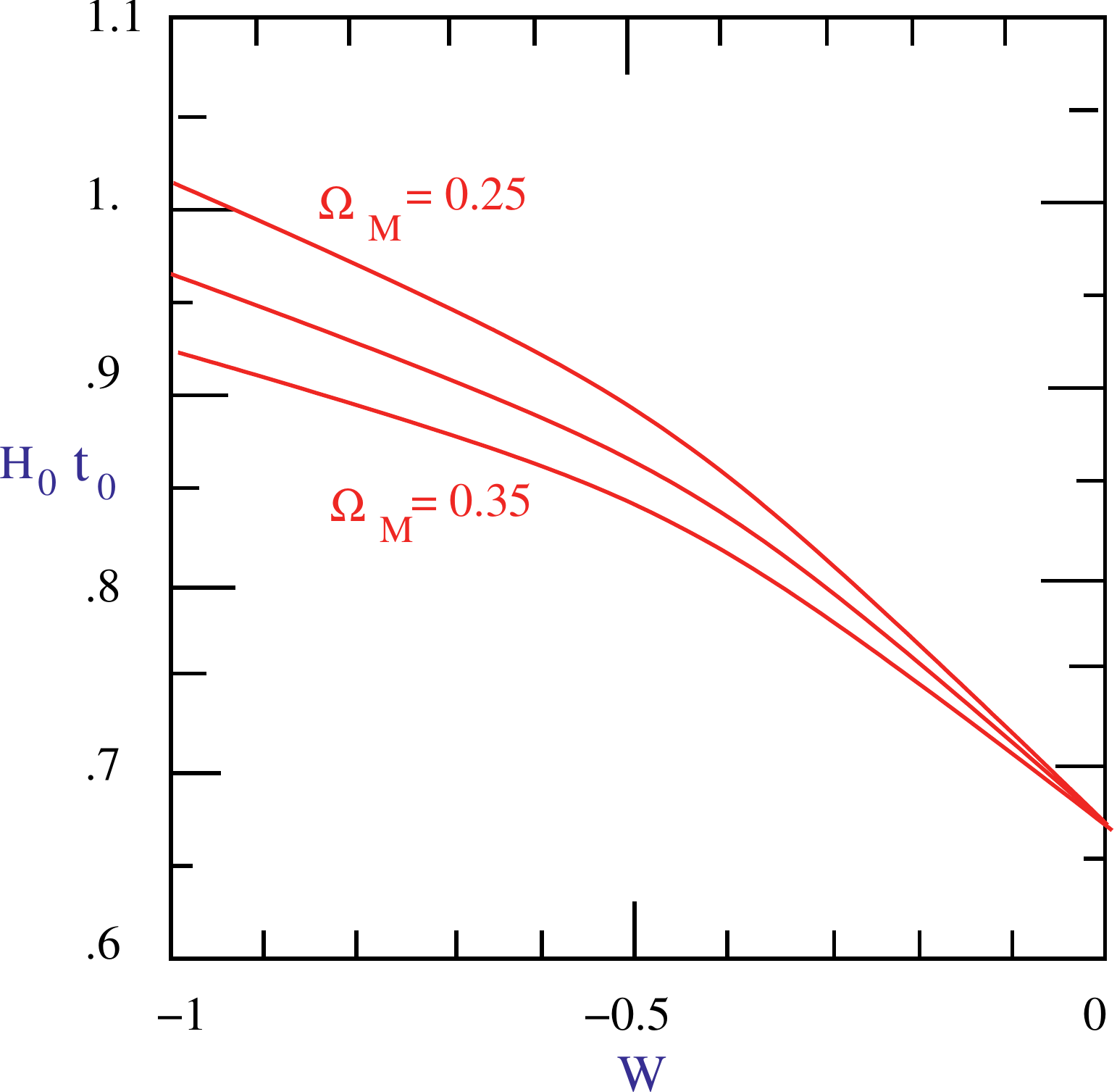}
\caption{
Evidence for the dark energy from Supernova data, left panel, and from the Universe age, right panel.
}
\label{fig:DarkEnergy}
\end{center}
\end{figure}

Supernova of type Ia have been  shown to be good standard candles. In measurements of luminosity distance to them the dark energy has been discovered. Below I illustrate this result in Fig.~\ref{fig:DarkEnergy}, left panel, using modern data and relation Eq.~(\ref{eq:Dmu}).  Blue curve corresponds to the Universe without dark energy, $\Omega_\Lambda = 0$, while violet curve corresponds to the best fit over dark energy which gives $\Omega_\Lambda = 0.7$. As a base cosmology which was subtracted I used the Universe with $\ddot{a} = 0$, but the subtraction  is not important here and is needed for the visualisation purposes only, to enhance separation of curves on a graph. 

In fact, prior to the dark energy discovery in Supernova data, scientists already suspected for a while that it  exists. 
Several hints existed, I illustrate the one derived from attempts to determine  the Universe age. 
During mater dominated expansion $a\propto t^{2/3}$. Therefore, for a matter dominated Universe without dark energy we would have $H_0\, t_0 = 2/3$.
However direct measurements of the Hubble constant gave at the time
$H_0 = 70 \pm 7\; {\rm km\; sec^{-1}\; Mpc^{-1}}$, while  the age was estimated (using ages of the oldest stars) as $t_0 = 13 \pm 1.5\; {\rm Gyr} $. Therefore, measurements producing 
$ H_0\, t_0 = 0.93 \pm 0.15$ were in contradiction with prediction for the matter dominated Universe.  

Let us see what happens if we add dark energy to a matter.
The age of the multicomponent universe can be found integrating $dt = a d\chi$ and using  $dz = - H d\chi $,  which for the universe age at redshift $z$ gives
$$
t(z) =   \int_z^{\infty} \frac{dz'}{(1+z')H(z')} \, . 
$$
In particular, if equation of state of dark energy corresponds to a vacuum, $w = -1$, and universe is spatially flat, $ \Omega_\Lambda + \Omega_M =1$, this gives
$$
H_0t_0 =  \frac{2}{3 \sqrt\Omega_\Lambda} \ln \left(\frac{1+ \sqrt\Omega_\Lambda}{\sqrt\Omega_M}\right)\, .
$$
Such a universe with $\Omega_\Lambda \simeq 0.7$ is a good fit to observations as opposed to a matter dominated universe, see Fig.~\ref{fig:DarkEnergy}, right panel and compare to a modern data in Table~\ref{tbl:cosmoparameters} which give $H_0t_0 \simeq 0.95$.

\section{Hot Big Bang}
\label{sec:HBB}

So far we have considered cosmography of the late Universe and found that the Universe should be filled with matter and dark energy.  However, the Universe should contain radiation also. Today its contribution is negligible, but in the early Universe it was dominant fraction. Indeed, energy density of radiation is fastest growing fraction when we look back in time, $a \rightarrow 0$, see Table~ \ref{tab:EqSate}. And the Universe was hotter as well in this limit, see Eq.~( \ref{eq:aT}). To reach such conclusions we have to assume also that the Universe was in thermal equilibrium in the past. But this is inevitable too since in a denser medium relaxation time is shorter. Universe was indeed in thermal equilibrium in the past, as we will shortly see. The concept of the Hot Universe is so natural and so inevitable, that it is hard to imagine nowadays that is was not widely accepted until relict radiation has beed observed.

\subsection{Cosmic Microwave Background Radiation}
\label{sec:CMBR}

The Universe is filled with radiation which is left-over from the Big Bang.
The name for this first light is Cosmic Microwave Background Radiation (CMBR).
Measurements of tiny fluctuations (anisotropy) in CMBR temperature give a
wealth of cosmological information and became a most powerful probe of
cosmology.

This radiation was predicted by Georgi Gamov in 1946, who estimated its
temperature to be $\sim 5\; K^{\circ}$. Gamov was trying to understand the
origin of chemical elements and their abundances. Most abundant, after
hydrogen, is helium, with its shear being $\sim 25\%$. One possibility which
Gamov considered was nucleo-synthesis of He out of H in stars. Dividing the
total integrated luminosity of the stars by the energy released in one
reaction, he estimated the number of produced He nuclei. This number was too
small in comparison with observations. Gamov assumed then that the oven where the
light elements were cooked-up was the hot early Universe itself. He calculated
abundances of elements successfully and found that the redshifted relic of
thermal radiation left over from this hot early epoch should correspond 
to  $\sim 5\; K^{\circ}$ at present. In one stroke G. Gamov founded two 
pillars (out of four) on which modern cosmology rests: CMBR and Big Bang
Nucleosynthesis (BBN). Hot Big Bang theory was born. 

Cosmic microwave background has been accidentally discovered by Penzias and Wilson
\cite{Penzias:1965wn} at Bell Labs in 1965 as the excess antenna temperature
which, within the limits of their observations, was isotropic, unpolarized,
and free from seasonal variations. A possible explanation for the observed
excess noise temperature was immediately given by Dicke, Peebles, Roll, and
Wilkinson and was published in a companion letter in the same issue
\cite{Dicke:1965}. Actually, they were preparing dedicated CMBR search experiment, but were
one month late. Penzias and Wilson measured the excess temperature as $\sim
3.5\pm 1\; K^{\circ}$. It is interesting to note that the first (unrecognized)
direct measurements of the CMB radiation was done by T. Shmaonov at Pulkovo in
1955, also as an excess noise while calibrating the RATAN antenna
\cite{Shmaonov:1957}. He published the temperature as $ (3.7 \pm 3.7)\;
K^{\circ}$. And even prior to this, in 1940, Andrew McKellar \cite{McKellar:1940} had
observed the population of excited rotational states of CN molecules in
interstellar absorption lines, concluding that it was consistent with being in
thermal equilibrium with a temperature of $ \approx 2.7 \; K^{\circ}$.  Its
significance was unappreciated and the result essentially forgotten.  Finally,
before the discovery, in 1964 Doroshkevich and Novikov in an unnoticed paper
emphasized \cite{Doroshkevich:1964} the detectability of a microwave blackbody
as a basic test of Gamov's Hot Big Bang model. 

The spectrum of CMBR is a perfect blackbody, with a temperature \cite{Fixsen}
\begin{equation}
T_0 = 2.7255 \pm 0.0006  \; K^{\circ},
\label{eq:TCMBR}
\end{equation} 
as measured by modern instruments. This corresponds to
410 photons per cubic centimeter or to the flux of 10 trillion photons per
second per squared centimeter. 

\subsection{Thermodynamics of the Universe}
\label{sec:ThermDynUniv}

There is no explanation to CMBR but the hot Big Bang. And since CMB is the radiation with black body spectrum, we know that the Universe was once in the thermal equilibrium.  It immediately follows from Eq.~(\ref{eq:aT}) that in the past the Universe was hotter since it was smaller. We can and should use thermodynamics describing the early Universe.

For particles  in  thermal equilibrium the phase-space distribution functions are:
\begin{equation}
f_i(k) =\frac{g_i}{(2\pi )^3} \, \frac{1}{\e_{~}^{(k_0-\mu_i)/T} \pm 1},
\label{eq:BoseFermiDist}
\end{equation} 
where $k_0$ is particle energy, $k_0 = \sqrt{\vec{k}^2+m_i^2}$,  $\mu$ is chemical potential  and the plus (minus) sign corresponds to fermions (bosons). Index $i$ refers to different particles species and $g_i$ is the number of their spin states, e.g. for photons, electrons and positrons $g_\gamma = g_{e^{-}} = g_{e^{+}} = 2$ correspondingly,  for neutrino and antineutrino $g_\nu = g_{\bar\nu} = 1$. All thermodynamical relations which we will need are derived using this function. In particular, number density of i-th particle species and their contribution into energy-momentum tensor are, correspondingly
\begin{eqnarray}
\label{eq:ND}
n_i &=& \int d^3 k  f_i(k),  \\
T_{\mu\nu} (i ) &=& \int d^3 k \frac{k_{\mu}k_{\nu}}{k_0} f_i(k).
\label{eq:EM}
\end{eqnarray}
Equation (\ref{eq:EM}) gives energy density as $\rho = T_{00}$, while  pressure is expressed through the trace over spatial part of the energy-momentum tensor, $p = -T_j^j/3$. To find overall energy density and pressure entering Friedmann equations one has to sum over all particle species, i.
 Entropy density is calculated as $ s = {(\rho + p - \mu n)}/{T}$.
Let us consider now important limits of these expressions.

\begin{figure}
\begin{center}
\includegraphics[width=0.49\textwidth]{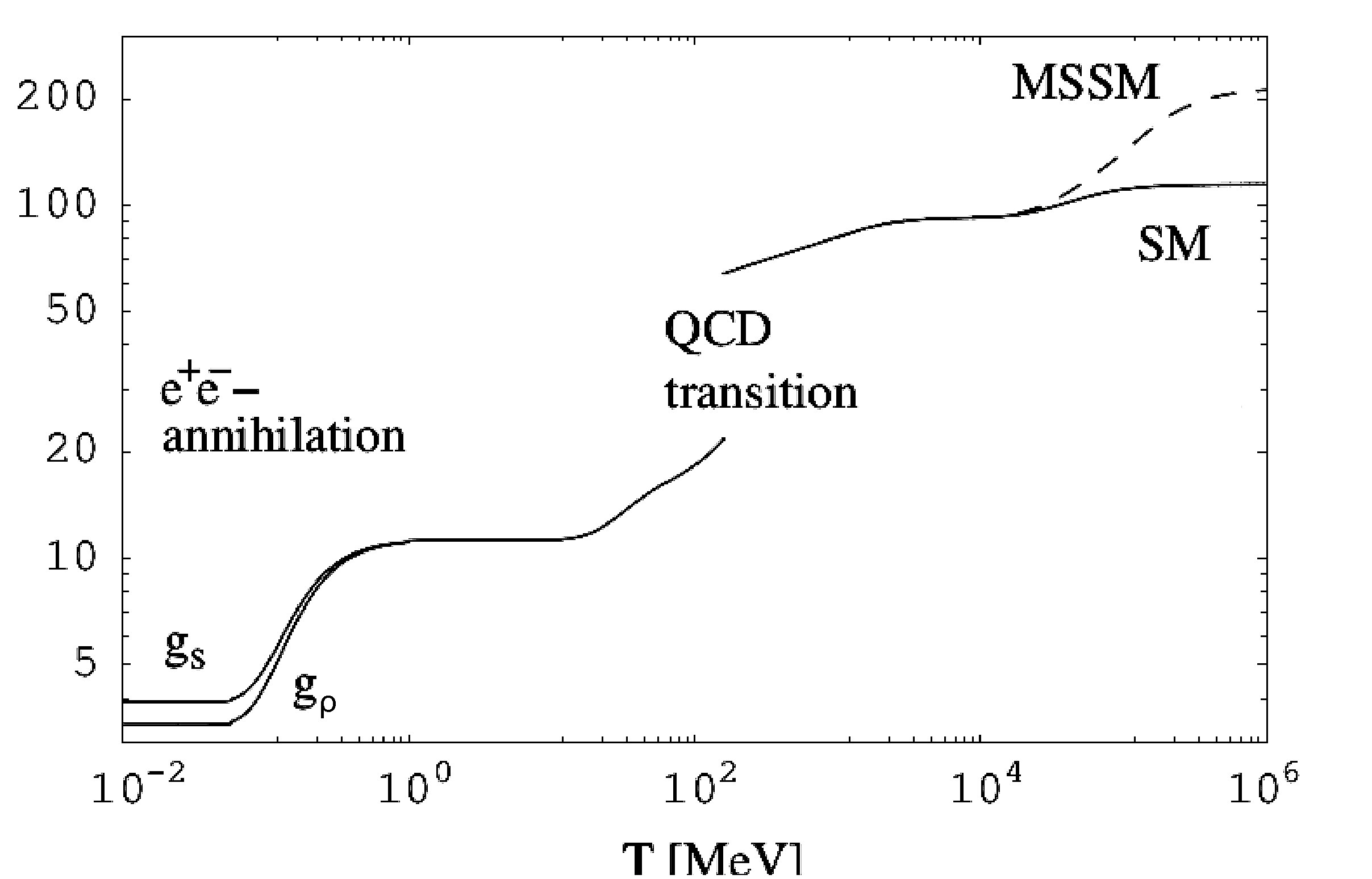}
\includegraphics[width=0.5\textwidth]{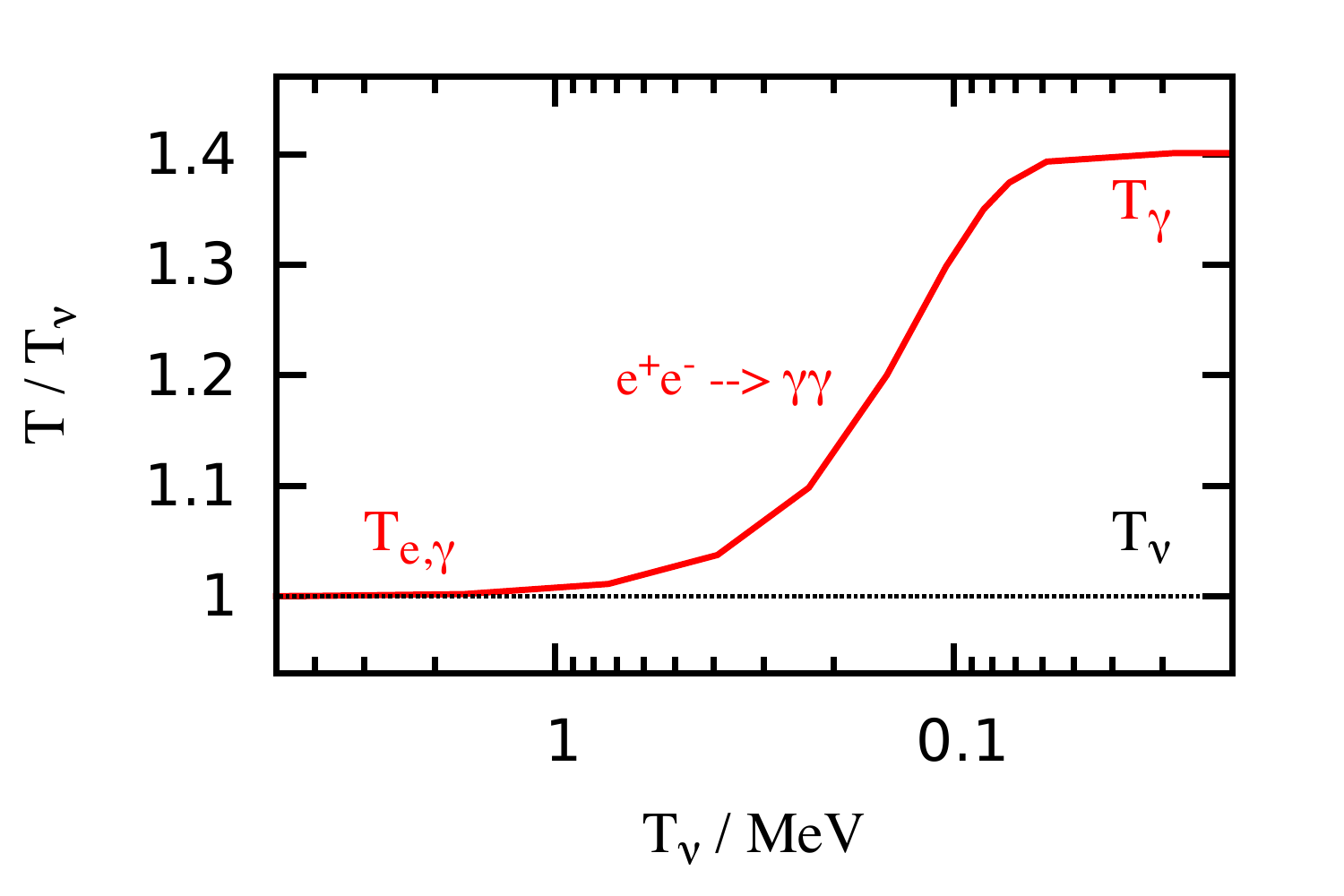}
\caption{Left panel. Number of relativistic degrees of freedom, $g_*$, as a 
function of temperature. Solid line -  the Standard Model case, dashed line - hypothetical behaviour in the Minimal Supersymmetric Standard Model (MSSM). Important events in the Universe evolutions are also indicated. Right panel. Increase of photon temperature over neutrino temperature during $e^+e^-$ annihilation.}
\label{fig:gT}
\end{center}
\end{figure}

1.  {\it Relativistic  particles.} First of all, for relativistic particles, regardless of particular form of $f(k)$, we have  $p = \rho/3$. In other words, this relation is valid even out of thermal equilibrium and simply follows from definitions since $k_0 = |\vec{k}|$ for $m = 0$. Further, for relativistic  plasma without chemical potentials, integrals in  Eqs.~(\ref{eq:ND}) and (\ref{eq:EM}) can be calculated analytically and are slightly different for bosons and fermions. Summing out over all particles we find
\begin{eqnarray}
\label{eq:Nrel}
n  &=& g_*' \, \frac{\zeta(3)}{\pi^2} \, T^3   \\
\label{eq:rhorel}
\rho  &=& g_*\,  \frac{\pi^2}{30}\; T^4 \\
\label{eq:Srel}
s  &=& g_*\,  \frac{2\pi^2}{45}\, T^3 
\end{eqnarray}
\small where $\zeta(3) \approx 1.2$ is Rieman zeta function and
\begin{eqnarray}\label{EntropyRadiation}
g_* '&=& {\footnotesize \sum_{{\rm bosons}} g_i +{3\over4}\sum_{{\rm fermions}} g_i }\,.\nonumber \\
g_* &=& {\footnotesize \sum_{{\rm bosons}} g_i +{7\over8}\sum_{{\rm fermions}} g_i}\,.\nonumber
\end{eqnarray}
In these expressions particles with $ m \ll T$ should be counted only,
i. e. ~$ g_*$ and $ g_*'$ are functions of the temperature. Temperature dependence of $g_*$ is  shown in Fig.~\ref{fig:gT}. Why it splits on $g_s$ and $g_\rho$ at $T \lsim \rm MeV$ will be explained later on.

2. {\it Non-relativistic  particles.} For non-relativistic particles all densities are exponentially suppressed  in thermal equilibrium and, again in the case without chemical potentials, we find
\begin{eqnarray}
\label{eq:NNonRel}
n_i&=&g_i \, \left( \frac{m_i T}{2\pi} \right)^{3/2}\,  \e^{-m_i/T}  \\
\rho_i  &=& m_i\, n_i + \frac{3}{2} T\, n_i \nonumber \\
p_i  &=& T\, n_i  \nonumber
\end{eqnarray}
Here expression (\ref{eq:NNonRel}) for $n_i$ is most important. In particular, it makes the basis for {\it Saha equation}, which will be used repeatedly throughout the lectures. This equation gives surviving amount of particles when they go out of equilibrium and will be used to discuss Big Bang nucleosynthesis, hydrogen recombination and abundance of thermally produced dark matter. 

\subsubsection{Cosmological density of neutrino}

In the expanding Universe particle concentrations, $n$, are in equilibrium as long as reaction rate is sufficiently high $\sigma n v > H$, where $\sigma$ is corresponding crossection. After that distributions do not change in a comoving volume, i.e. "freeze-out".
Weak interaction rate for neutrino $\sigma^{~}_W n \sim G_F^2 T^2\cdot T^3$ matches  expansion
rate, $H \sim T^2/M_{\rm Pl}$, when $G_F^2 M^{~}_{\rm Pl} T^3 \approx 1$. We conclude that neutrino are in thermal equilibrium at temperatures $T \gg 1$ MeV and decouple from the rest of plasma
at lower temperatures. Therefore, Standard Model neutrinos, which have
small masses, decouple when they are still relativistic.  The number density
of neutrino at this time is given by Eq.~(\ref{eq:Nrel}) with $g_*' = 2$.
Below this temperature, neutrinos are no longer in thermal equilibrium with
the rest of the plasma, and their temperature simply decreases as $T \propto
1/a$. However, the cosmological background of photons is heated up by the
$e^{+}e^{-}$ annihilations shortly after neutrino "freeze-out". Let us find a relation between $T_\nu$ and
$T_\gamma$, which will also give the relation between $n_\nu$ and $n_\gamma$.


Recall that entropy in the comoving volume conserves, 
$g_*\; T^{3} = {\rm const}$. Before annihilation $g_* = g_\gamma + g_e\cdot (7/8) =
2 + 4\cdot (7/8) = 11/2$.  After annihilation $g_* = g_\gamma = 2$. (Neutrinos are decoupled already and do not participate in these relations.) Since
before annihilation $T_\nu = T_\gamma$ the condition $g_*\; T^{3} = {\rm const}$ gives for the neutrino temperature after positron annihilation 
\begin{equation}
T_\nu = \({4}/{11}\)^{1/3} T_\gamma.
\label{eq:NuTemp}
\end{equation}
This can be compared to the result of numerical integration of corresponding Boltzmann equations which is shown in Fig.~\ref{fig:gT}, right panel.  
Present day photon temperature $ T_\gamma = 2.7255~ K$, therefore, present day neutrino temperature is $ T_\nu = 1.9454~ K$.
For the number density of one  flavour of left-handed neutrino and antineutrino we find
$n_\nu = {3} n_\gamma /11 = 115\; {\rm cm}^{-3}$.
Here we have used Eq.~(\ref{eq:Nrel})  and $g_\gamma = g_\nu = 2$. Right-handed
neutrino, even if exist and light, are not in thermal equilibrium  and are excited in small amounts, see Ref.~\cite{Dolgov:2002wy} and Section \ref{sec:RHnu_asDM} about "sterile" neutrino as dark matter.  

Assume that by now neutrino became non-relativistic, i.e. their masses are larger than the present day temperature. In this case, neutrino energy
density is given by $ \rho_\nu = \sum_i m_{\nu i}\, n_{\nu i}$. Since it has
to be smaller than $\Omega_m\, \rho_c $, we already have the constraint
$\sum_i m_{\nu i} < 93 \; \Omega_m h^{2}\; {\rm eV} \approx 10 \; {\rm eV}$.  Modern cosmological constraints on neutrino masses are almost two orders of magnitude stronger actually (see later in the lectures).

Now, using Eqs.~(\ref{eq:rhorel}) and (\ref{eq:NuTemp}) we find that after "freeze-out" and $e^+e^-$ annihilation (but at $T > m_\nu$) the  cosmological radiation background is parametrized as
\begin{equation}
\rho_r = \rho_\gamma + \rho_\nu =  \frac{\pi^2}{15}\; T_\gamma^4 \, \left[1+\frac{7}{8}\left(\frac{4}{11}\right)^{4/3}N_{eff}\right] .
\label{eq:rhorad}
\end{equation}
At face value $N_{eff}$ stands here for the number of active neutrino flavours and should be equal to three.
But actually according to conventions used by cosmologists, $ N_{eff} \neq 3$, neither it is integer.  The reasons are as follows:
\begin{itemize}
\item  When  $e^+e^-$ annihilate, neutrino are not decoupled completely yet since neutrino "freeze-out" temperature $\approx 1$ MeV is too close to $e^+e^-$ annihilation temperature. 
This leads to slight neutrino heating with 
distorted distribution (\ref{eq:BoseFermiDist})
and to somewhat larger neutrino energy density $\rho_\nu$ \cite{Dolgov:1992qg}, which in applications is parametrised simply as larger $N$, to account only for the increase of  $\rho_\nu$,
$$N_{eff} = 3.046.$$ 
\item   There can be other contributions into radiation,  e.g.  light sterile neutrinos, Goldstones, or some other very light particles. These contributions are called  "dark radiation". They are also included into  $N_{eff} $ and "dark radiation" is searched for in modern data as a signal that $N_{eff} > 3.046$.
\end{itemize}
Therefore, $ N_{eff}$  is another important cosmological parameter, potentially signalling new physics.

Now we can also understand why $g_*$ splits on $g_s$ and $g_\rho$ at $T \lsim \rm MeV$. At these temperatures, radiation consists of two fractions with different temperatures each, gas of photons and gas of neutrino.
Therefore, when writing Eqs.~(\ref{eq:rhorel})  and (\ref{eq:Srel}) we have two options. We could have two terms in each of these equations, one for photons, another for neutrino, each term would have different temperatures. Or we can do the same way as in Eq.~(\ref{eq:rhorad}), including ratio of temperatures into $g_*$ instead. And this latter approach has been decided to be more convenient by cosmologists.  Since temperature enters in different powers to energy and entropy densities, we have splitting of $g_*$ on $g_s$ and $g_\rho$. Asymptotic values of these functions, which can be used at $T < 100$~keV,  are shown below, assuming there is no dark radiation
\begin{equation}
\label{eq:grho}
g_\rho (0) = 2 + 3.046\;   \left(\frac{7}{4}\right)\,   \left(\frac{4}{11}\right)^{{4}/{3}}  = 3.38,
\end{equation}
\begin{equation}
\label{eq:gs}
g_s (0) = 2 + 3.046\,  \left(\frac{7}{4}\right)\,  \left(\frac{4}{11}\right) ^{ {~}~{~}} = 3.94.
\end{equation}

\paragraph{Matter-radiation equality}

Radiation energy density scales with expansion as $\rho \propto a^{-4}$, while matter energy density scales as $\rho \propto a^{-3} $. It follows that the Universe was radiation dominated at the early stages of the evolution.
Let us find now at which cosmological redshift and temperature the very important event happens: namely, when the energy density of radiation becomes equal to the energy density of matter. Using
Eq.~(\ref{eq:rhorad})  with  $N_{eff} = 3.046$ and present day photon temperature, Eq.~(\ref{eq:TCMBR}), we obtain  
$\rho_r = 4.41 \times 10^{-10} \rm\; {GeV}\; {cm^{-3}}$ for the current radiation energy density.
Recall now the value of critical density,
$\rho_c = 1.05\times 10^{-5}\; h^2\;\rm{GeV}\;{cm^{-3}} $, Eq.~[\ref{eq:CritRho}), ~to get~
$ \Omega_r = 4.2 \times 10^{-5} \, h^{-2} $.
Since radiation scales as $\rho_r = \rho_c \Omega_r (1+z)^{4}$ while matter as $\rho_m = \rho_c \Omega_m (1+z)^{3}$, we find for the redshift of matter-radiation equality
$$1+z_{\rm eq} = \frac{\Omega_m}{\Omega_r} = 3400,$$
and for the corresponding temperature $T_{\rm eq} = 0.8$ eV. Deriving this  I used values for $\Omega_m$ and $h$ from Table~\ref{tbl:cosmoparameters}. Keeping those as free parameters we have $T_{eq} = 5.6\, \Omega_m\, h^2\; {\rm eV}$.

At higher temperatures the Universe was radiation dominated and its expansion was governed by the following Hubble parameter
\begin{equation}
H = \sqrt{\frac{8\pi G \rho_{r}}{3}} = \sqrt{\frac{8\pi^3 g_* T^4}{90\, M^2_{\rm Pl}}}  \simeq 1.66\, \sqrt{g_*}\, \frac{T^2}{M_{\rm Pl}} .
\label{eq:HT}
\end{equation}
Since during radiation dominated stage $H =1/2t$, we obtain   the Universe age (in seconds) as a function of temperature
$$t  ({\rm s}) = \frac{2.42}{\sqrt{g_*}} \left(\frac{\rm MeV}{T}\right)^2 . $$
Stretching this time-temperature relation to equality temperature, and using expression (\ref{eq:grho}) for $g_*$, we find that at equality the Universe was 65 thousand years old.

\subsection{Last scattering of light}

Baryonic matter is  ionized at temperatures higher than the hydrogen ionization energy $E_{\rm ion} =13.6$ eV
 and photons are in thermal equilibrium with primordial plasma. They cannot propagate  large distances and the plasma is not transparent. With expansion the Universe cools down. At some point protons and electrons of primordial plasma recombined into neutral hydrogen and the Universe became transparent for radiation. This happens  when the mean free path of photons becomes comparable to the size of the Universe at that time. Corresponding temperature is called "last scattering".  After that photons are travelling without being affected by scattering.
We see this light as Cosmic Microwave Background Radiation (CMBR). More precisely, the CMBR comes from the surface of the last scattering. We cannot see past this surface. Let us determine here when the last scattering had occurred in the early Universe.

Fraction of ionized hydrogen as function of temperature can be described by the Saha equation. It is derived by simply making ratios of number densities, Eq.~(\ref{eq:NNonRel}), of different fractions in question in thermal equilibrium. For the case of hydrogen recombination
\begin{equation} 
\frac{n_e\,
n_p}{n^{~}_H} = \( \frac{m_e T}{2\pi}\)^{3/2}\e ^{-E_{\rm ion}/T}.
\label{Saha}
\end{equation}
Here $n_e,~ n_p$ and $n^{~}_H$ are the number densities of electrons, protons,
and neutral hydrogen respectively.
Plasma is electrically neutral, i.e. $n_e = n_p$. To find closed relation
for the fraction of ionized atoms, $X\equiv n_p/(n_p+n^{~}_H) = n_p/n^{~}_B$,
we need the relation between the baryon number density, $n^{~}_B$, and
temperature. This relation can be parameterized with the help of an important
cosmological parameter called {\bf baryon asymmetry} 
\begin{equation} 
\eta =
\frac{n^{~}_B}{n_\gamma} =\frac{n_p + n^{~}_H}{n_\gamma} = (6.1 \pm 0.05)\times
10^{-10} \; ,
\label{BAU}
\end{equation}
where ${n_\gamma}$ is the number density of photons, Eq.~(\ref{eq:Nrel}).
Baryon asymmetry can be estimated by an order of magnitude by simply counting the
number of baryons, or comparing element 
abundances predicted by the theory  of Big Bang Nucleosynthesis to observations. Those are not 
most precise methods, though; the value presented in Eq.~(\ref{BAU}) was
obtained from fitting the spectrum of CMBR fluctuations, see below. Nowadays,
this is the most precise baryometer. 

Defining recombination as the temperature when $X=0.1$, we find
$T_{\rm rec} \approx 0.3~{\rm eV}$.
The Universe became transparent for radiation when the mean free path of
photons became comparable to the size of the Universe at that time. Photons
scatter mainly on electrons and we find that the Universe became transparent
when
\begin{equation}
(\sigma_{\gamma e}\, n_e)^{-1} ~\sim~ t  \; .
\label{TranspCond}
\end{equation} 
Here, $\sigma_{\gamma e} = 8\pi \alpha^{2}/3m_e^{2}$ is the Compton
cross-section.  For the temperature of last scattering we find $T_{\rm ls}
\approx 0.26\; {\rm eV}$. Taking the ratio to the current CMBR temeperature
we find  $z_{\rm ls} \approx 1000$.

CMBR is the oldest light in the Universe. When registering it, we are looking
directly at the deepest past we can, using photons. These photons had traveled
the longest distances without being affected by scattering, and geometrically
came out almost from the universe Horizon.  Therefore the CMBR gives us a snapshot of the
baby Universe at  the time of last scattering.

\section{ CMB power spectrum: tool of Precision Cosmology}
\label{sec:CMBR_tool}

\begin{figure}
\begin{center}
\parbox{4cm}{\vspace{-4.7cm}\hspace{-3.0cm}
\includegraphics[width=0.49\textwidth]{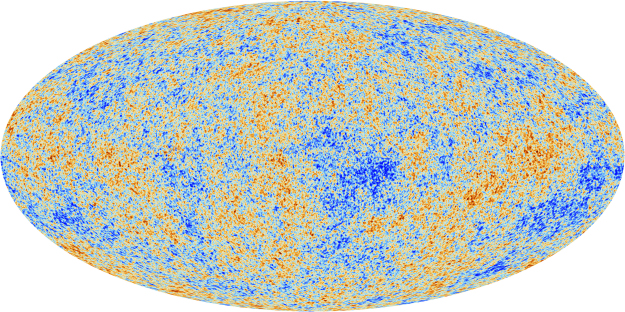}}
\hspace{3.7cm}
\includegraphics[width=0.5\textwidth]{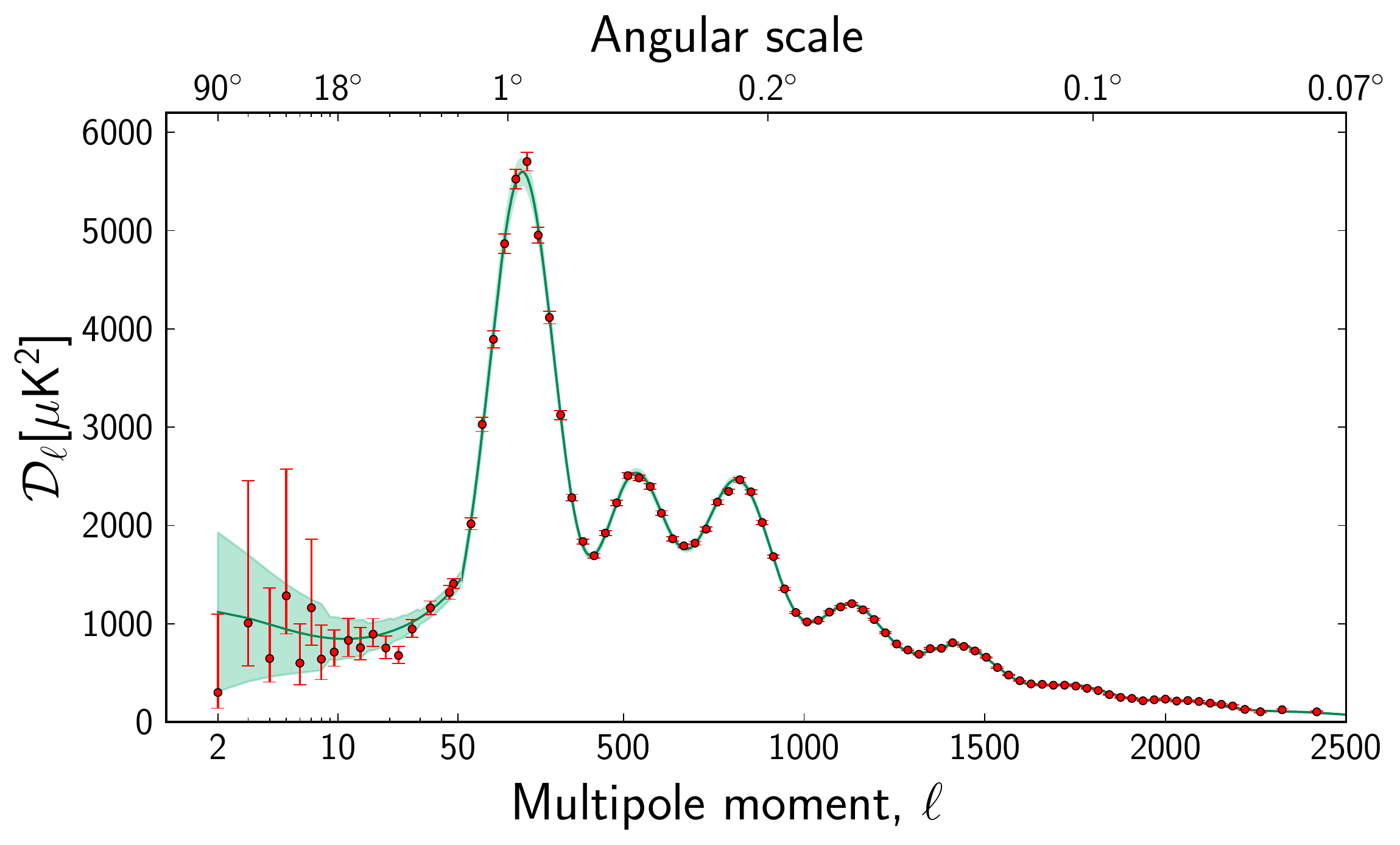}
\caption{Sky map of primordial temperature fluctuations in Galactic coordinates (left panel) and temperature power spectrum (right panel) as measured by Planck  space observatory~\cite{Ade:2015xua}.}
\label{fig:CMB}
\end{center}
\end{figure}

The temperature of CMBR is slightly different in different patches of the sky - to 1 part in 100,000. These temperature deviations are shown in the sky map Fig.~\ref{fig:CMB}, left panel.  Measurements of these tiny fluctuations (anisotropy) in CMBR temperature give us a wealth of cosmological information at an unprecedented level of precision and became a most powerful probe of cosmology. The functional form of the CMBR power spectrum is very sensitive to both the
various cosmological parameters and to the shape, strength and nature of primordial fluctuations. 
This spectrum is shown in Fig.~\ref{fig:CMB}, right panel. In fact, the values of cosmological parameters listed in Table I largely came out from fitting   model predictions to data as in this figure.

The temperature anisotropy, $T({\bf n})$, as a function of viewing direction
vector ${\bf n}$,  as shown in Fig.~\ref{fig:CMB}, left panel, is naturally expanded in a  basis of spherical harmonic,
$Y_{lm}$
\begin{equation}
T({\bf n}) = \sum\limits_{l,m} a_{lm} Y_{lm}({\bf n})\; .
\label{T-SpericHarmDec}
\end{equation}
Coefficients $a_{lm}$ in this decomposition define the angular power
spectrum, $C_l$
\begin{equation}
C_l={\frac{1}{2l+1}}\sum \limits_m \vert a_{lm} \vert ^2\; .
\label{T-PowSpectr}
\end{equation}
Assuming random phases, the r.m.s. temperature fluctuation assosiated with the angular
scale $l$ can be found as 
\begin{equation} \Delta T_l = \sqrt{C_l\; l(l+1) / 2\pi} \equiv \sqrt{D_l}.
\label{DeltaT}
\end{equation}
Spectrum, ${D_l}$, as measured by Planck collaboration, is shown in
Fig.~\ref{fig:CMB}, right panel. In fact, it was realized already
right after the discovery of CMBR,  that fluctuations in its
temperature should have fundamental significance as a reflection of the seed
perturbations which grew into galaxies and clusters. In a pure baryonic
Universe it was expected that the level of fluctuations should be of the order
$\delta T /T \sim 10^{-2}-10^{-3}$. Mesurements of the CMBR anisotropy with
ever-increasing accuracy have begun. Once the temperature fluctuations were
shown to be less than one part in a thousand, it became clear that baryonic
density fluctuations did not have time to evolve into the nonlinear structures
visible today. A gravitationally dominant dark matter component was invoked.
For explanations why it is necessary, see Section~\ref{sec:DM}.
Eventually, fluctuations were detected \cite{Smoot:1992td} at the level of $ \delta T / T ~\sim
~10^{-5}$, consistent with the structure formation in Cold
Dark Matter models with the Harrison-Zel'dovich spectrum of primordial
perturbations motivated by cosmological Inflation, see Section~\ref{sec:inflation} and Appendices.   

The foundations of the theory of CMBR anisotropy were set out by Sachs \&
Wolfe \cite{Sachs:1967er}, Silk \cite{Silk:1968kq}, Peebles \& Yu
\cite{Peebles:1970ag}, Syunyaev \& Zel'Dovich \cite{Syunyaev:1970}.  The
measured spectrum of CMBR power has a characteristic shape of multiple
peaks. Positions of these peaks and their relative amplitudes are sensitive to
many cosmological parameters in a non-trivial way. Fitting the data to model
predictions gives very accurate values for many of these parameters (though
there are some degeneracies between deferent sets). Numerical calculations for
different models were done already in Ref. \cite{Doroshkevich:1978}, and power
spectra exhibiting acoustic peaks (similar to those in
Fig.~\ref{fig:CMB}, right panel) were presented. It was realized, in particular,
that positions of the peaks are shifted with respect to each other for
adiabatic and isentropic primordial fluctuations.

\begin{figure}
\begin{center}
\includegraphics[width=0.58\textwidth]{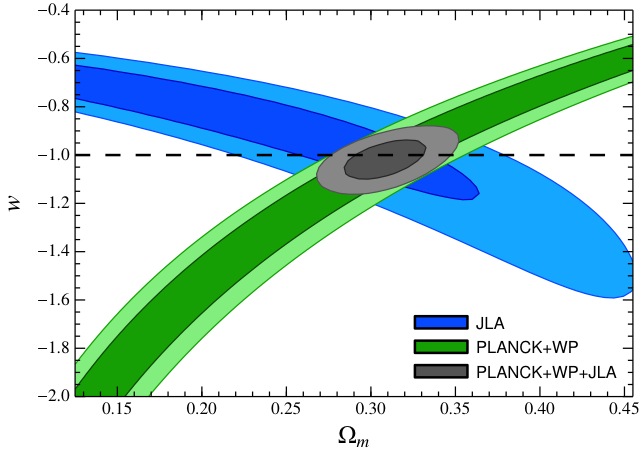}
\hspace{.5cm}
\includegraphics[width=0.377\textwidth]{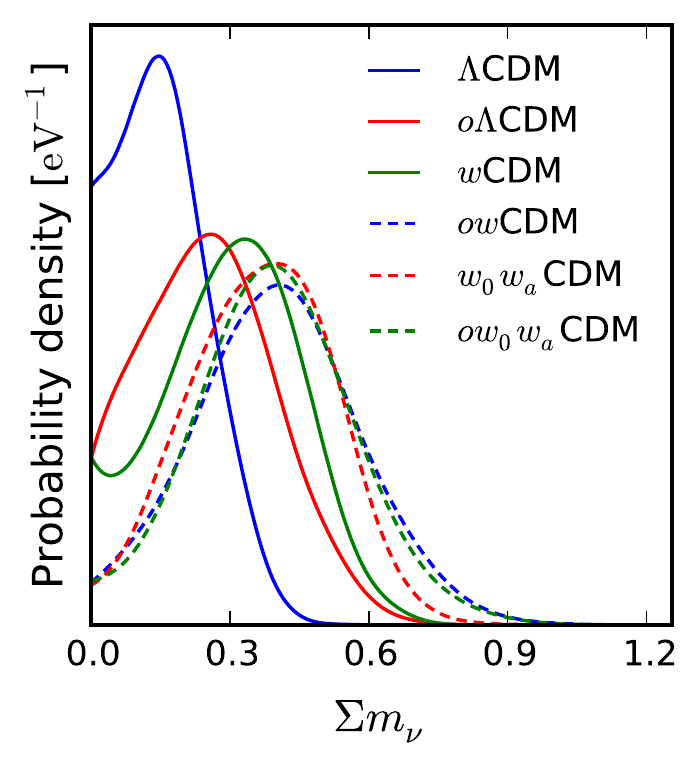}
\caption{Left panel: Combination of SN Ia (blue contours) and Planck data (green contours) tell us that the equation of state of the dark energy is consistent with that of a vacuum, $w = -1$, from Ref~\cite{Betoule:2014frx}. Rigt panel: Constraints on neutrino mass from combined Planck and BAO data,  from Ref~\cite{Pellejero-Ibanez:2016ypj}.}
\label{fig:CMB_JLA_BAO}
\end{center}
\end{figure}

To improve significantly constraints on some cosmological parameters a combination of CMBR with other data is needed. For example, combining Planck data with Supernova data we find that the Dark energy equation of state is close to a vacuum, $w=-1.02 \pm 0.06$~\cite{Betoule:2014frx}, while each of these sets alone would give weak constraints, see  Fig.~\ref{fig:CMB_JLA_BAO}, left panel. Combination of Planck data with data on correlation properties of galaxy clustering, namely data on Baryon Acoustic Oscillations (BAO)  tells us that the Universe is spatially flat,  
 $1-\sum\Omega_i = 0.000 \pm 0.005$, see Ref.~\cite{Ade:2015xua}. 
That same data set improves many other constraints.  An example of constraints on the sum  of neutrino masses in this joint data set is shown in Fig.~\ref{fig:CMB_JLA_BAO}, right panel~\cite{Pellejero-Ibanez:2016ypj}. Solid blue line corresponds to the case of $\Lambda$CDM model, which means zero spatial curvature and  $w = -1$. Other curves on this figure correspond to a Dark entry models with equation of state different from that of a vacuum. For the $\Lambda$CDM the constraint on the neutrino masses is $\sum m_\nu \rm  < 0.22 ~\rm eV$ with positive $2\sigma$ detection if Dark energy is more complicated substance than vacuum. However, $\Lambda$CDM is a good model and is consistent with all cosmological data to date.

Therefore  $\Lambda$CDM can be safely assumed, and then other cosmological parameters can be determined quite well from the CMBR data alone. For example, parameters presented in Table~\ref{tbl:cosmoparameters} (except spatial curvature) were obtained from the CMBR data alone assuming $\Lambda$CDM model. Note that the constraint on $\Omega_\Lambda$ from the supernova luminosity distance relations, Section \ref{sec:SNLumDist}, I  also obtained  assuming the $\Lambda$CDM model. Otherwise from the SN data alone we would only know for sure that the Universe expansion is accelerating, but the fraction of dark energy, $\Omega_{DE}$, would depend upon assumed equation of state $w$, as it is shown  by blue contours in Fig.~\ref{fig:CMB_JLA_BAO},  left panel.

\subsection{Acoustic oscillations}

As we could see already, large amount of cosmological information is encoded in the functional form of the CMBR power spectrum. To get feeling of physics which is behind,
let us give a qualitative picture of why the CMBR power spectrum has a
specific shape of a sequence of peaks, and explain how it depends on the
values of particular cosmological parameters. Insight, sufficient for the
purposes of these lectures, can be gained with the idealization of a perfect
radiation fluid. In complete treatment, one has to follow the evolution of
coupled radiation and metric fluctuations, i.e. to solve the linearized
Einstein equations. However, essential physics of radiation (or matter)
fluctuations can be extracted without going into the tedious algebra of
General Relativity. It is sufficient to consider the energy-momentum
conservation, Eq.~(\ref{relativ-energy-conservation}). To solve for metric
perturbations, full treatment based on Einstein equations,
Eq.~(\ref{Einstein-equations}), is needed of course. We will not do that here,
but simply quote results for the evolution of the gravitational potentials
(coincident in some important cases with the solutions for the Newtonian
potentials).

Perturbations of the ideal radiation fluid, $p = \rho/3$, can be separated
into perturbations of its temperature, velocity and gravitational potential.
In the general-relativistic treatment gravitational potential appears as a
fractional perturbation of the scale factor in the perturbed metric 
\begin{equation} ds^{2}
= a^{2}(\eta) \, [(1+2\Psi)d\eta^{2} - (1-2\Phi)dx^{i}dx_j] \; .
\label{metric-perturbed}
\end{equation} 
Two equations contained in the energy-momentum conservation,
$T^{\mu\nu}_{\,\,\,\,\,\,\,\,\,;\nu}=0$ (i.e. temporal $\mu=0$ and spatial
$\mu=i$ parts of this equation), written in metric (\ref{metric-perturbed}),
can be combined to exclude the velocity perturbations. The resulting
expression is simple
\begin{equation} 
\ddot{\theta}_k + \frac{k^{2}}{3} \theta_k = - \frac{k^{2}}{3}
\Phi_k + \ddot{\Phi}_k \; .
\label{eqMotion-radFluid}
\end{equation}
Note that this equation is the exact result for a pure radiation fluid.  Here,
$\theta_k$ are Fourier amplitudes of $\delta T/T$ with wavenumber $k$, and
$\Phi_k$ is a Fourier transform of gravitational potential.  Analysis of
solutions of the Einstein equations for $\Phi$ shows that $\Phi_k $ do not depend 
upon time in two important cases:
\begin{enumerate}
\item For superhorizon scales, which are defined as $k\eta \ll 1$.
\item For all scales in the case of matter dominated expansion, $p=0$.
\end{enumerate}
In these situations the last term in the r.h.s. of
Eq.~(\ref{eqMotion-radFluid}), namely, $ \ddot{\Phi}_k$, can be neglected. 
The Einstein equations also restrict the initial conditions for fluctuations.
For the adiabatic mode in the limit  $k\eta \ll 1$ one finds 
\begin{equation}
\delta_{0k} = - 2\Phi_{0k} \; ,
\label{GravPot-InCond}
\end{equation}
where $\delta \equiv \delta \rho/\rho$, and subscript $0$ refers to the initial values. 
The adiabatic mode is defined as a
perturbation in the total energy density. For the one component fluid, which
we consider here, only the adiabatic mode can exist.  Note that fractional
perturbation of the scale factor in metric (\ref{metric-perturbed}),
$a(\eta,{\bf x}) = a(\eta) +\delta a(\eta,{\bf x}) \equiv a(\eta) (1-\Phi)$,
can be expressed as perturbation of spatial curvature, see
Eq. (\ref{Spatial-curvature}). Therefore, adiabatic perturbations are also
called curvature perturbations.  Let us re-write Eq. (\ref{GravPot-InCond}) for
temperature perturbations:
\begin{itemize}
\item Radiation domination,  $\delta  = 4\, \delta T/T $, and we find 
\begin{equation}
\theta_{0k} = - \frac{\Phi_{0k}}{2}\; .
\label{theta-phi-rd}
\end{equation}
\item Matter domination,  $\delta  = 3\, \delta T/T $, and we find 
\begin{equation}
\theta_{0k} = - \frac{2\Phi_{0k}}{3}\; .
\label{theta-phi-md}
\end{equation}
\end{itemize}
Recall now that in the limit $k\eta \ll 1$ the gravitational potentail is
time-independent, $\Phi = {\rm const}$. Therefore,
Eq.~(\ref{eqMotion-radFluid}) has to be supplemented by the following initial
conditions in the case of the adiabatic mode:
\begin{equation}
\theta_{0k} \neq 0, \hspace{1cm} \dot{\theta}_{0k}= 0\; .
\label{inCond-adiabatic}
\end{equation}

\paragraph{Temperature fluctuations on largest scales.} 
Let us consider the modes which had entered the horizon after matter-radiation
equality, $ k \eta_{\rm eq} < 1$.  For those modes, $\dot{\Phi} = 0 $ all the
way from initial moments till present, and the solutions of
Eq.~(\ref{eqMotion-radFluid}) with adiabatic inital conditions is

\begin{equation}
\theta_k + \Phi_k =
(\theta_{0k} + \Phi_{0k})\, \cos \(\frac{k\eta}{\sqrt{3}}\)\; .
\label{radFluid-solution}
\end{equation}
As gravity tries to compress the fluid, the radiation pressure resists. As in everyday physics, 
this  leads to acoustic oscillations. But here, it is important that oscillations are synchronized.
All modes have the same phase regardless of $k$. This is a consequence of
$\dot{\theta}_{0k} = 0$, which is valid for all $k$. At the last scattering, the
universe becomes transparent for the radiation and we see a snapshot of these
oscillations at $\eta = \eta^{}_{\rm ls}$.

To get its way to the observer, the radiation has to climb out of the
gravitational wells, $\Phi$, which are formed at the last scattering surface.
Therefore the observed temperature fluctuations are $\theta_{\rm obs}~=
\theta(\eta^{}_{\rm ls} ) + \Phi$, or
\begin{equation}
\theta_{k,\rm obs} = \frac{1}{3}
\Phi_{0k}\, \cos \(\frac{k\eta^{}_{\rm ls}}{\sqrt{3}}\) \; ,
\label{Sachs-Wolfe}
\end{equation}
where we have used Eq.~(\ref{theta-phi-md}), which relates initial values of
$\theta$ and $\Phi$.  Note that overdense regions correspond to cold spots in
the temperature map on the sky, since the gravitational potential is
negative. This is  famous Sachs-Wolfe effect \cite{Sachs:1967er}.

\paragraph{Acoustic peaks in CMBR.}

Modes caught in the extrema of their oscillation, $k_n \eta^{}_{\rm
ls}/\sqrt{3} = n \pi$, will have enhanced fluctuations, yielding a fundamental
scale, or frequency, related to the universe sound horizon, $s_* \equiv
\eta^{}_{\rm ls}/\sqrt{3}$.  By using a simple geometrical projection, this
becomes an angular scale on the observed sky. In a spatially flat Universe,
the position of the first peak corresponds to $l_1 \approx 200$, see below.
Both minima and maxima of the cosine in Eq.~(\ref{Sachs-Wolfe}) give peaks in
the CMBR power spectrum, which follow a harmonic relationship, $k_n = {n
\pi}/s^{}_{*}$,~ see Fig.~\ref{fig:CMB}, right panel.

The amplitudes of the acoustic peaks are recovered correctly after the
following effects are taken into account: 
\begin{enumerate}
\item Baryon loading. The effect of added baryons is exactly the same for the oscillator equation
Eq.~(\ref{radFluid-solution}), as if we had increased the mass of a load
connected to a spring, which oscillates in a constant gravitational field and with the 
starting point on the top of an uncompressed coil at rest. The addition of baryons makes
a deeper compressional phase, and therefore increases every other peak in the
CMBR power spectrum. (First, third, fifth, $\dots$) The CMBR power spectrum is
a precise baryometer.
\item Time-dependence of $\Phi$ after horizon crossing in radiation dominated
universe. Gravitational potentials are not constant, but decay inside the horizon during
radiation domination. This decay drives the oscillations: it is timed in such a way that 
compressed fluid has no gravitational force to fight with, when the fluid turns
around.  Therefore, the amplitudes of the acoustic peaks increase as the cold
dark matter fraction decreases, which allows to measure $\Omega_m$.
\item Dissipation. This leads to a dumping of higher order peaks in the CMBR power spectrum.
\end{enumerate}

\paragraph{Position of the first peak.}
Position of the first peak is determined by the angular size of the sound
horizon at last scattering. Let us calculate here a similar quantity: the
causal horizon (which is larger by a factor of $\sqrt{3}$ in comparison with
the sound horizon).  The comoving distance traveled by light, $ds^{2} = 0$,
from the ``Big Bang'' to redshift z is determined by a relation similar to
Eq.~(\ref{eq:chi_z}), but with different integration limits
\begin{equation} 
\chi (z) = \int_z^{\infty}
\frac{dz'}{H(z')} \; ,
\label{com-fromInfty-toZ}
\end{equation} 
where $H(z)$ is given by Eq.~(\ref{eq:H_z}).  One has to
integrate this relation with a complete set of $\Omega_i$. However,
from the last scattering to $z \sim 1$,
the Universe was matter dominated.  Therefore, the causal horizon in a matter
dominated Universe 
$
\chi (z) = {2}/{H_0\sqrt{1+z}}
$
should give a reasonable first approximation to the true value of integral in Eq~(\ref{com-fromInfty-toZ}).
Consider now two light rays registered at $z=0$ which were separated by a comoving
distance $\chi(z_{\rm ls}^{~})$ at the moment of emission. Since both propagate in
the metric $ ds^{2} = a^{2}(d\eta^{2} - d\chi^{2} - \chi^{2} d\theta^{2}) = 0$,
we find for the angular size of horizon at last scattering
\begin{equation}
\theta_h =\frac{\chi (z_{\rm ls})}{\chi (0)- \chi (z_{\rm ls})}  \approx
\frac{1}{\sqrt{1+z_{\rm ls}}} = \sqrt{\frac{T_0}{T_{\rm ls}}} \approx
1.7^\circ\; .
\label{angSize-hor-ls}
\end{equation}
Note that this is an approximate relation since we had neglected the dark energy contribution into 
expansion of the Universe at late stages. To get sound horizon we have to divide Eq.~(\ref{angSize-hor-ls}) by $\sqrt{3}$.
Observationally, the sound horizon angular scale is tightly constrained by Planck from the position of the first peak:
$ \theta_* = 0.59648^\circ \pm 0.00018^\circ$ \cite{Ade:2015xua}. This is important direct observable, which can be used to set  constraints on cosmological parameters entering Eq.~(\ref{angSize-hor-ls}). 

\paragraph{Horizon problem.}
\label{sec:HorProblem}

Relation (\ref{angSize-hor-ls}) tells us that regions separated by more than $\;>\;2^\circ$ on the map of microwave sky, Fig.~\ref{fig:CMB}, have not been in the causal
contact prior to the last scattering in the standard Friedmann cosmology.   We should see $10^{4}$ causally disconnected regions at the surface of last scattering. Temperature could vary wildly from point to point which are further away from each other than  $2^\circ$. Yet, CMBR temperature  is the same 
to better than~ $10^{-4}$ accuracy all over the sky. Observations tell us that
all sky regions were somehow synchronized according to the adiabatic initial
conditions, Eq.~(\ref{inCond-adiabatic}), with only small initial
perturbations present, $\Phi_i \ll 1$. This constitutes the so-called
``Horizon problem`` of standard cosmology. In Section~\ref{sec:inflation} we
will see how this problem is solved in frameworks of inflationary cosmology.

\section{Inflationary Cosmology}
\label{sec:inflation}

In frameworks of ``classical'' cosmology and assuming no fine-tuning, one
concludes that a typical universe should have had Plankian size, live Plankian
time and contain just a few particles.  This conclusion is based on the
observation that Fridmann equations contain single dimension-full parameter
$M_{\rm Pl} \sim 10^{19}$ GeV, while dimensionless parameters naturally are
expected to be of order unity. Yet, the observable Universe contains $10^{90}$
particles in it and had survived $10^{65}$ Plankian times. Where does it all
came from?  In other words, why is the Universe so big, flat ($\Omega_0
\approx 1$) and old ($t > 10^{10}$ years), homogeneous and isotropic ($\delta
T/T \sim 10^{-5}$), why does it contain so much entropy ($S>10^{90}$) and does
not contain unwanted relics like magnetic monopoles?  These puzzles of
classical cosmology were solved with the invention of Inflation
\cite{Starobinsky:1979ty,Starobinsky:1980te,Guth:1981zm,Linde:1982mu,Albrecht:1982wi,
Linde:1983gd}. All these questions are related to the initial conditions and
one can simply postulate them.  The beauty of Inflation is that it generates
these unnatural initial conditions of Big Bang, while the pre-existing state
(which can be arbitrary to a large extent) is forgotten. Moreover, with development, 
Inflationary theory delivered unplanned bonuses. Not only does the Universe become clean and
homogeneous during inflation, but also the tiny perturbations necessary for
the genesis of galaxies are created with the correct magnitude and
spectrum. Below we consider the basics of Inflationary cosmology.

\subsection{Big Bang puzzles and Inflationary solutions}

By definition, Inflation is a period in the Universe evolution when $\ddot{a} > 0$ .
Using the second Friedmann equation, Eq.~(\ref{eq:fridman2}), we find that the
inflationary stage is realized when $p < - \rho /3$. In particular, if $p = -
\rho$ the energy density remains constant during expansion in accord with the
first law of thermodynamics, Eq.~(\ref{FirstLawTherm}), and the
physical volume expands exponentially fast,
$a(t) =\e^{Ht}$, see~ Eq.~(\ref{eq:fridman1}). Let us see now how the condition $\ddot{a} > 0$ during some early stage solves problems of classical cosmology.

\paragraph{Horizon problem}

In Section \ref{sec:HorProblem} we have found that  the angular size of
horizon at the moment of last scattering is $\approx 2^\circ$, see
Eq.~(\ref{angSize-hor-ls}), which tells us that we observe $10^{4}$ causally
disconnected regions at the surface of last scattering. The question then arises,
why is the Universe so homogeneous at large scales?

This problem can be solved if during some period of time  the Universe expansion was
sufficiently fast. To find quantitative requirement, let us consider a power low
for the Universe expansion, $a(t) \propto t^\gamma$. The physical size of a given
patch grows in proportion to the scale factor, $R_{\rm P} \propto \;a(t)$.
On the other hand, Eq.~(\ref{com-fromInfty-toZ}) tells us that  the physical size of a causally 
connected region (horizon) grows in proportion with  time, $R_{\rm H} = a \chi = t/(1-\gamma)$. 

The exponent $\gamma$ depends upon the equation of state,
$\gamma = 1/2$ for radiation and $\gamma = 2/3$ for the matter dominated
expansion. In any case, for the ``classical'' Friedmann Universe $\gamma < 1$
and the horizon expands faster than volume.  Take the largest visible patch
today. It follows that in the past its physical size should have been larger
than the horizon size (since they are equal today) and therefore
this patch should have contained many casually disconnected regions.  
On the other hand, if during some period of evolution $\gamma > 1$, the
whole visible Universe could have been inflated from one (``small'') causally connected region.
In such cosmology, any given patch in the Universe passes the boundary of causally connected region twice. First when it is inflated and becomes bigger than horizon, and second, when the inflationary stage changes to "Big Bang" and casually connected region at some future point in time becomes larger than this given patch. Note that $\gamma > 1$ means $\ddot a > 0$.

\paragraph{Curvature problem}

The first Friedmann equation~(\ref{eq:fridman1}) can be re-written as
\begin{equation} 
k = a^{2}\(\frac{8\pi G}{3}\, \rho - H^{2}\) = 
a^{2}H^{2}\,\(\Omega -1\) = \dot{a}^{2}\,\(\Omega -1\).
\label{Friedmann-Eq1a}
\end{equation} 
Since $k$ is a constant, we immediately see the problem: during matter or radiation dominated
stages $\dot{a}^{2}$ decreases (this happens for any expansion
stage with $\ddot{a} < 0$), and $ \Omega$ is driven away from
unity.  However, at present we observe $\Omega ~\approx~ 1$.  Therefore, initially the Universe
has to  be extremely  fine-tuned, say at the epoch of
nucleosynthesis, when temperature was $T\sim 1$ MeV, one should have $|\Omega(t_{\rm
NS}) -1| < 10^{-15}$, and even stronger tuning is required at earlier epochs.
A possible solution is obvious: accelerated expansion $\ddot{a} > 0$ increases
$\dot{a}$ and therefore drives $\Omega(t)$ to unity prior to radiation dominated stage.  A robust prediction of inflationary cosmology is a flat Universe, $\Omega =1$.

\paragraph{The problem of Entropy}

As we know already, the energy of a vacuum stays constant despite
the expansion.  In this way, room for matter full of energy could have been
created.  If there is mechanism to convert vacuum energy into particles and radiation at some
later stage, then  the observed huge entropy will be
created and the problem of entropy will be solved. Potentially, this mechanism works for any inflationary scenario,
since the product $\rho a^{3}$ is guaranteed to grow whenever $\ddot{a} >
0$. However, the important question is whether a graceful exit out of the
inflationary stage and successful reheating is indeed possible. In practice, this
issue has killed a number of inflationary models. Remarkably, the original
model by A. Guth \cite{Guth:1981zm} had being ruled out precisely on these
grounds \cite{Guth:1981uk}.

\paragraph{For how long the inflationary stage  should last?}

Inflation has to continue for a sufficiently long time for the problems of
horizon, curvature and entropy to be solved. All these requirements  give roughly the same
condition on the number of  ``e-foldings'' of inflation
\cite{Guth:1981zm} and we present here a (simplified) derivation based on
entropy. 
Multiplying the current temperature in the universe by its visible size we find
$T_0 a_0 \chi_0 \sim 10^{30}$, where $\chi_0$ is the comoving size of the present
horizon. We also want the whole visible universe
to be inflated out of a single causally connected patch. A given wave mode is in vacuum state when its wavelength is smaller than the size of Hubble parameter during inflation and becomes frozen as a classical fluctuation when it becomes larger. This is just a consequence of the quantum field theory in a universe expanding with acceleration, $\ddot a > 0$, see Appendices. Therefore, important inflationary period, which can be linked to observations,  is from the moment when the patch corresponding to the whole visible Universe goes out of the Hubble scale and to the moment when inflation ends, i.e. from the moment  $a_i\chi_0 = H^{-1}$ until $a_r = \e^{N_*} a_i$, where  the number of e-foldings, $N_* \equiv H \Delta t$, parametrises duration of this inflationary period. At the end of inflation the vacuum energy goes to radiation with temperature $T_r$  which is related to the present day temperature as $T_r a_r =   T_0 a_0$, see
Eq.~(\ref{eq:aT}), and we  neglect here the change in the number of
relativistic degrees of freedom from  $T_r $ to  $T_0 $. This gives 
\begin{equation} 
\frac{T_r}{H}\; \e^{N_*} \approx 10^{30} \; .
\label{e-folds}
\end{equation} 
Number of e-foldings $N_*$ depends upon reheating temperature.
In popular models of Inflation the ratio ${T_r}/{H_i}$ is within a couple
orders of magnitude from unity, and we find $50 \lsim N_* \lsim 60$. 
I stress again that $N_*$ is not the duration of inflation. The latter cannot be smaller than $N_*$, but inflation can last longer of  course, and then our Universe is homogeneous to scales much much larger than its visible part today.

\subsection{Models of Inflation}

Consider energy-momentum tensor $T_{\mu\nu}$ for a scalar field {$\varphi$}
\begin{eqnarray}
T_{\mu\nu}=\partial_{\mu}\varphi\, \partial_{\nu}\varphi - 
g_{\mu\nu}\, {\cal L},
\label{Tmunu-scalarfield}
\end{eqnarray} 
with the Lagrangian 
\begin{eqnarray}
{\cal L}=\partial_{\mu}\varphi\, \partial^{\mu}\varphi - V(\varphi)\; .
\label{Lagrangian-scalarfield}
\end{eqnarray} 
In a state when all derivatives of {$\varphi$} are zero, the stress-energy
tensor of a scalar field simplifies to
$T_{\mu}^\nu=V(\varphi)\,\delta_{\mu}^\nu$. This corresponds to a vacuum
state. Indeed, comparing with Eq.~(\ref{Tmunu-ideal}), we find $V = \rho = -p$. A large number of inflationary models exists where {$\varphi \approx$ const} during some period of evolution and  {vacuum}-like state is imitated. Such field is called inflaton.

1. {\it False vacuum inflation.} 
Conceptually simple and easily understandable scenario  was suggested by {A.~Guth} \cite{Guth:1981zm}. Consider potential $V(\phi)$ which has a local minimum with
a non-zero energy density separated from the true ground state by a
potential barrier.  A universe which happened to be trapped in the
meta-stable minimum will stay there for a while (since such a state can decay
only via subbarrier tunneling) and expansion of the universe will diminish all
field gradients. Then the Universe enters a vacuum state and Inflation starts.  Subsequent phase transition into the true minimum
ends inflationionary stage and creates the radiation phase.
Today the model of Guth and its variants based on potential barriers is 
good for illustration purposes only. It did not stand up to
observations since inhomogeneities which are created during the
phase transition into the radiation phase are too large \cite{Guth:1981uk}. But the model gives 
easily understandable answer to the  frequently asked question:
how can it be that the energy density stays constant despite the expansion?

2. {\it Chaotic inflation.} Andrei Linde was first to realize that things work in the simplest possible
setup \cite{Linde:1983gd}.  Consider potential
\begin{equation}
V(\phi)=\frac{1}{2}m_\phi^2\phi^2.
\label{chaotic}
\end{equation}
Field equation  in an expanding Universe and for the homogeneous mode 
is
$
\ddot{\phi}+3H\dot{\phi}+m_\phi^2\phi=0.
$
If {$H \gg m_\phi $}, the ``friction'' $3H\dot{\phi}$ dominates and the field does not
move (almost). Therefore, time derivatives in $\; T_{\mu\nu}$ can be neglected, and
inflation starts (in a sufficiently homogeneous patch of the Universe). A
Hubble parameter in this case is determined by the potential energy, $H
\approx m_\phi {\phi}/{M_{\rm Pl}}$, and we see that inflation starts if the
initial field value happened to satisfy ${\phi > M_{\rm Pl}}$. During
inflationary stage the field slowly rolls down the potential hill. This motion
is very important in the theory of structure creation, see Appendices. Inflation ends when
${\phi \sim M_{\rm Pl}}$. At this time, field oscillations start around the
potential minimum and later decay into radiation. In this way all matter content has been 
likely created in our Universe. In general, this model generalises to arbitrary monomial $V(\phi)\propto \phi^\alpha$ as field potential at large $\phi$.

3.  {\it $R^2$ - inflation.} Historically, this is  the first model of inflation. It was  invented by A. Starobinsky~\cite{Starobinsky:1979ty,Starobinsky:1980te}.
Einstein-Hilbert action, leading to Einstein equations (\ref{Einstein-equations}) should be modified inevitably in quantum field theory on a curved space-time. In particular, counter-terms proportional to  the squares of different curvature tensors should be added to cancel divergences. Starobinsky considered the simplest form of extended gravitational Lagrangian
\begin{equation}
{\it L} = \frac{M_{\rm Pl}^2}{2} R + \beta R^2,
\label{eq:Starob}
\end{equation}
where $R$ - scalar curvature and $\beta$ - some dimensionless constant. Universe inflates in this model. It can be understood as follows. After conformal rotation this model is equivalent to the usual Einstein gravity plus a scalar field with potential
$$
V(\phi) = \Lambda^4\left(1 - \e^{-\sqrt{2/3}\, \phi/M_{\rm Pl}} \right)^2.
$$
This potential has very flat plateau at $\phi > M_{\rm Pl}$, and with such initial value of $\phi$ the Universe will inflate. The Universe will be heated up in the same way is in chaotic inflation after $\phi$ will slowly reach $M_{\rm Pl}$.

\subsection{Unified theory of Creation}
\label{sec:creation}

During Inflation  the Universe was in a vacuum-like state.  We
have to figure out how this ``vacuum'' had been turned into the matter we observe
around us, and how primordial fluctuations which gave rise to galaxies were
created. Solution to all these problems can be understood in a single  unified
approach. Basically, everything reduces to a problem of particle creation in
a time-dependent classical background. On top of every ``vacuum'' there are
fluctuations of all quantum fields which are present in a given model. This
bath of virtual quanta is indestructible, and even Inflation cannot get rid of
it. Being small, fluctuations  obey an oscillator equation
\begin{equation}
\ddot{u}_k ~+~ [k^2 + m^2_{\rm eff} (\eta)]\; u_k = 0 \, ,
\label{ModEq}
\end{equation}
here $u_k$ are amplitudes of fluctuating fields in Fourier space.  Effective
mass becomes time dependent through the coupling to time-dependent
background. Because $m_{\rm eff}$ is time dependent, it is not possible to
keep fluctuations in a vacuum.  If oscillators with
momentum $k$ happened to be in the vacuum at one time, they will not be in the vacuum at a
latter time because positive and negative frequency solutions mix, see Appendices.
Several remarks are in order.
\begin{itemize}
\item Eq.~(\ref{ModEq}) is valid for all particle species.      
\item The equation looks that simple in a conformal reference frame $ds^2 =
      a(\eta)^2\;(d\eta^2-dx^2)$. (And a ``dot'' means
      derivative with respect to $\eta$.)
\item Of particular interest are ripples of space-time itself: curvature
      fluctuations (scalar fluctuations of the metric) and gravity waves
      (tensor fluctuations of the metric).
\item Effective mass $m_{\rm eff}$ can be non-zero even for massless fields.
      Gravitational waves give the simplest example \cite{Grishchuk:1975ny},
      with $m^2_{\rm eff} = - {\ddot{a}}/{a}$. The effective mass for curvature
      fluctuations has a similar structure $m^2_{\rm eff} = - {\ddot{z}}/{z}$,
      but with $a$ being replaced by $z \equiv a\dot{\phi}/H$, see Refs.
      \cite{Lukash:1980iv,Sasaki:1986hm,Mukhanov:1988jd,Mukhanov:1992me}.
\item For a scalar field which does not couple to the inflation, the effective mass  is given by Eq.~(\ref{eq-motion-scalar-conformal}). For conformally coupled, but massive scalar it reduces to $m_{\rm eff} = m_0 \, a(\eta)$.
\end{itemize}
Note that creation in Inflationary theory is possible because nature is not
conformally-invariant. Otherwise, $m_{\rm eff}$ would be time-independent and vacuum
would remain vacuum forever.  There are two important instances of time varying
classical background in cosmology:
expansion of space-time, ~$a(\eta )$,
and motion of the inflaton field, ~$\phi (\eta )$.
Both can be operational separately or together at any epoch  of creation:
\begin{itemize}
\item During inflation. This is when superhorizon size perturbations of metric are created, which
give seeds for   the formation of galaxies and Large Scale Structure in general.
\item After inflation while the inflaton oscillates. This is when  matter itself is
created out of energy generated from the vacuum. 
\end{itemize}
There are several primary observables which can be calculated out of $u_k$ and
further used for calculation of quantities of interest. Most useful are:
\begin{itemize}
\item The particle occupation numbers, $n_k$. Integration over $d^3k$
gives the particle number density.
\item The power spectrum of field fluctuations, $P(k) \equiv u_k^*
u_k$. Integration over $d^3k$ gives the field variance.
\end{itemize}
Depending on physical situation, only one or the other may have
sense. The particle number in a comoving volume is useful because it 
 is adiabatic invariant on sub-horizon  scales (or when $m>H$) and
allows to calculate  the amount of created matter and abundances of various relics, e.g. dark matter.
But it has no meaning at super-horizon scales when $m < H$.  Then the power spectrum
of field fluctuations  is used which allows to calculate density perturbations and gravitational waves generated
during inflation.
Necessary details of such calculation are given in Appendices. 

\subsubsection{Testing Inflationionary predictions}

Typically, the spectrum of curvature perturbations generated during inflations has a form $P_\zeta (k) = A_s k^{n_s}$, where  $A_s $ and ${n_s}$ are constants (i.e. weakly depend upon scale $k$). Similarly, for gravitational waves  $P_T (k) = A_T k^{n_T}$.
To the first approximation, the Hubble parameter $H$ during inflation is
constant. Then, power spectra do not depend on $k$
and $n_S=1$, $n_T=0$. This case is called the Harrison-Zel'dovich spectrum 
\cite{Harrison:1970fb,Zeldovich:1972ij} of primordial perturbations which has been suggested on general grounds before inflationary theory was invented.  
However, in reality, $H$ is changing and these constants take different, model dependent values. Nevertheless, there is model independent relation between the
slope of tensor perturbations and the ratio of power in tensor to curvature
modes 
\begin{equation}
r \equiv \frac{P_{T}(k)}{P_\zeta(k)} = -8 n_T  \; .
\label{consistency-relation}
\end{equation}
This is called the {\it consistency relation} to which (simple) inflationary
models should obey. It will be robust and ultimate test of inflationary theory when imprint of gravitational waves in a CMBR will be discovered.

\begin{figure}
\begin{center}
\includegraphics[width=0.75\textwidth]{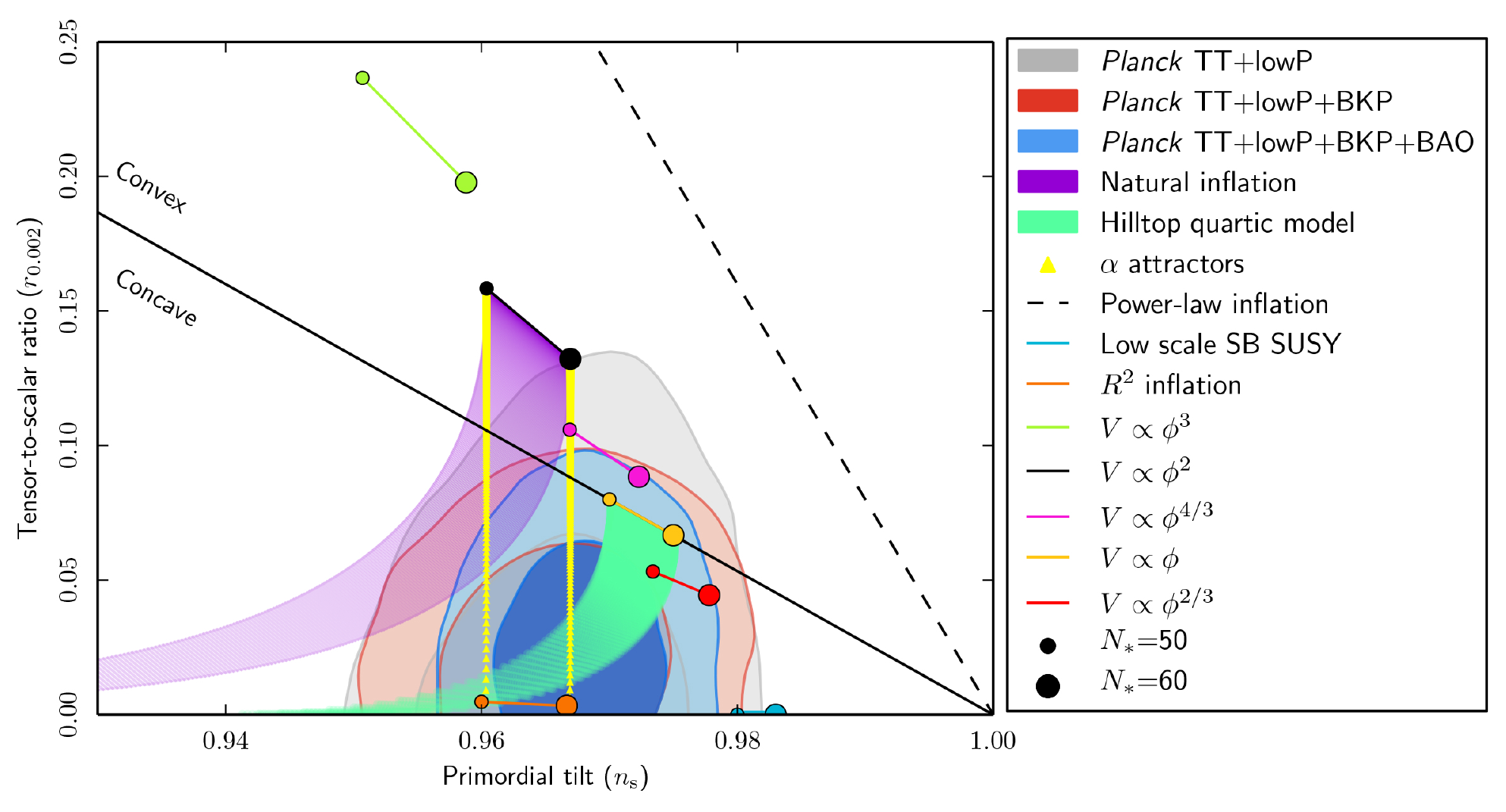}
\caption{Constraints on inflationary models in the
$(n_s,r)$ plane, from Ref.~\cite{Ade:2015lrj}.  Coloured line segments with circles at ends correspond to predictions of different inflationary models  with different inflation potentials. Within each segment $N_*$ varies in the interval $50 \leq N_* \leq 60$, see Eq. (\ref{e-folds}).  }
\label{fig:SDSS_MAP-inflation}
\end{center}
\end{figure}

However, both $A_s$ and $n_s$ are measured, extracted from CMBR observations and can be compared to model predictions.
The most recent constraints in the $(n_s,r)$ plane, obtained by Planck collaboration~\cite{Ade:2015lrj}  are presented in Fig.~\ref{fig:SDSS_MAP-inflation}. We see that in chaotic inflationary model, Eq.~(\ref{chaotic}), the gravitational waves would have been already discovered by Planck, and this model is ruled out nowdays. Best runner is $R^2$ model of A. Starobinsky, Eq.~(\ref{eq:Starob}), which is a perfect fit. However, observation of the imprint of gravity waves  in this model will be very difficult task, if possible at all.

To summarise, all predictions of Inflationary cosmology, which could have being tested so
far, have being confirmed. In particular, the Universe is spatially flat
(within experimental errors), see Table I. The
primordial perturbations are of superhorizon size and adiabatic. The spectral
index is close to the Harrison-Zeldovich case, see
Fig.~\ref{fig:SDSS_MAP-inflation}. Crucial test of
inflationary paradigm would be detection of gravity waves and verification of
the consistency relation.

\section{Dark Matter}
\label{sec:DM}

We have seen already in Section~\ref{sec:CMBR_tool} that 
CMBR observations accurately measure the nature and spectrum of the primordial fluctuations, the geometry of the Universe, its present expansion rate and its composition, see also Table~\ref{tbl:cosmoparameters}, which is based on these observations.
In particular, these measurements tell us  that in addition to baryonic matter there should be also dark matter which so far had been seen only through its gravitational influence.  This "sterility" leaves open the possibility that in fact we should look for modification of gravity, not for dark matter, in order to explain the missing mass problem. While both possibilities are exciting and beyond contemporary physics, a successful modified gravity theory  was not constructed yet. Therefore, I will not discuss numerous  attempts  and various models of modified gravity here, instead I'll  just give two original references, the early one \cite{Milgrom:1983ca}, and the most recent  one  \cite{Verlinde:2016toy}. It is difficult to construct such a theory for several reasons. In particular, the evidence for missing mass exists at various scales and epochs while modification should explain everything. Contrary to that, e.g. simple variants of MOND \cite{Milgrom:1983ca} do explain the "missing mass" on galactic scales  without invoking dark matter, but fail to explain other evidence. Moreover, MOND is phenomenological, non-relativistic prescription, not a theory. Therefore, other  cosmological tests, beyond CMBR,  are  also important. Below we consider cosmological observations that are independent of the CMB but also point to the existence of non-baryonic dark matter. At the end of the section I briefly discuss some popular models of dark matter and present status of dark matter searches in corresponding models.

\subsection{Dark Matter: the evidence}
\label{sec:DMevidence}

Missing mass is seen on all cosmological scales. In particular,  it reveals itself as
\vspace{0.3cm}

\begin{itemize}
  \item Flat rotational curves in galaxies;
  \item Gravitational potential which confines galaxies
    and hot gas in clusters;
  \item Gravitational lenses in clusters;
  \item Gravitational potential which allows structure
    formation from tiny primeval perturbations;
  \item Gravitational potential which creates CMBR anisotropies.
\end{itemize}
In this subsection I shortly review this overwhelming evidence for the unseen, but gravitating mass.

\subsubsection{Dark Matter in Galaxies}

\begin{figure}
\begin{center}
\includegraphics[width=0.49\textwidth]{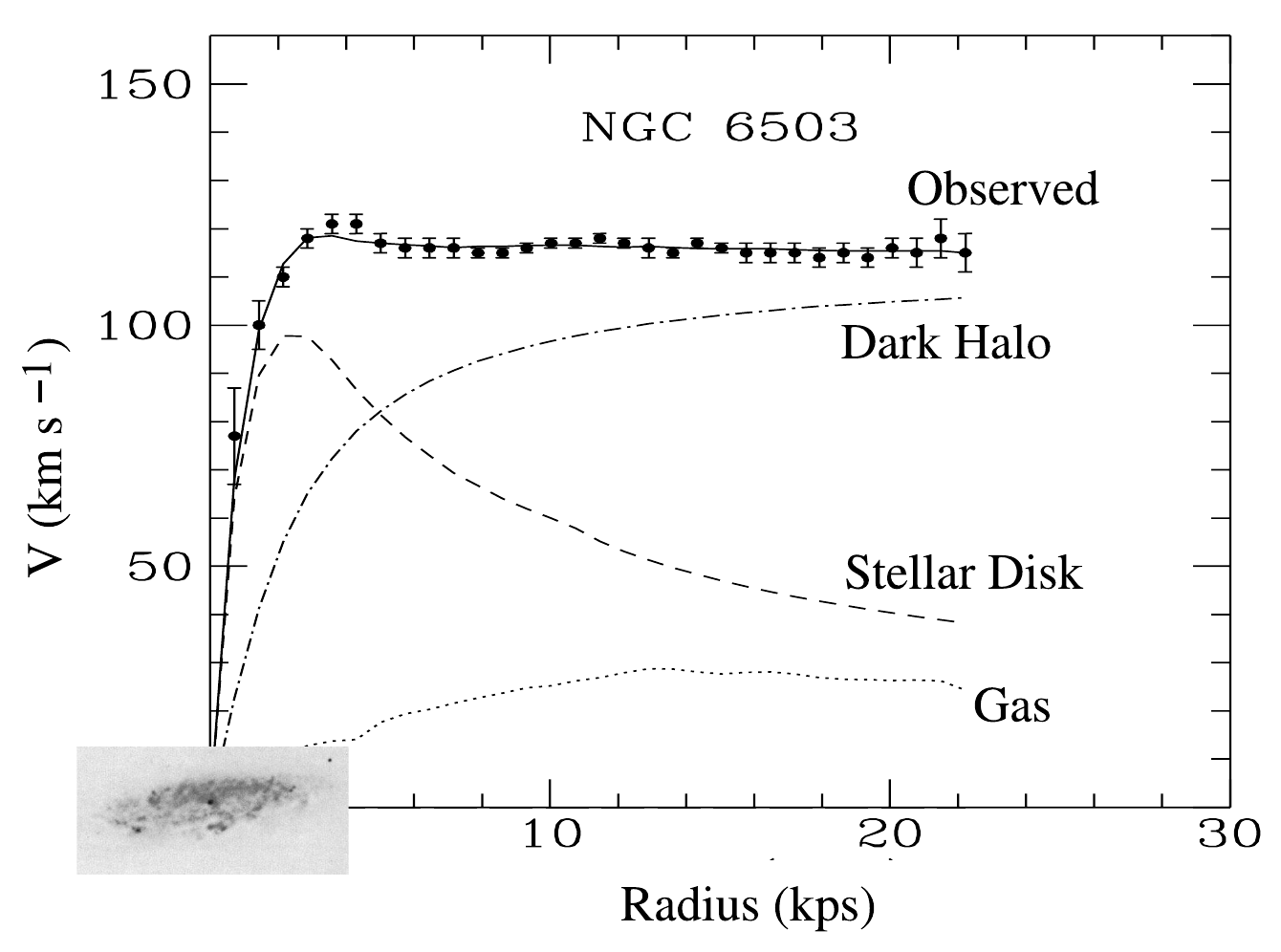}
\hspace{0.1cm}
\includegraphics[width=0.49\textwidth]{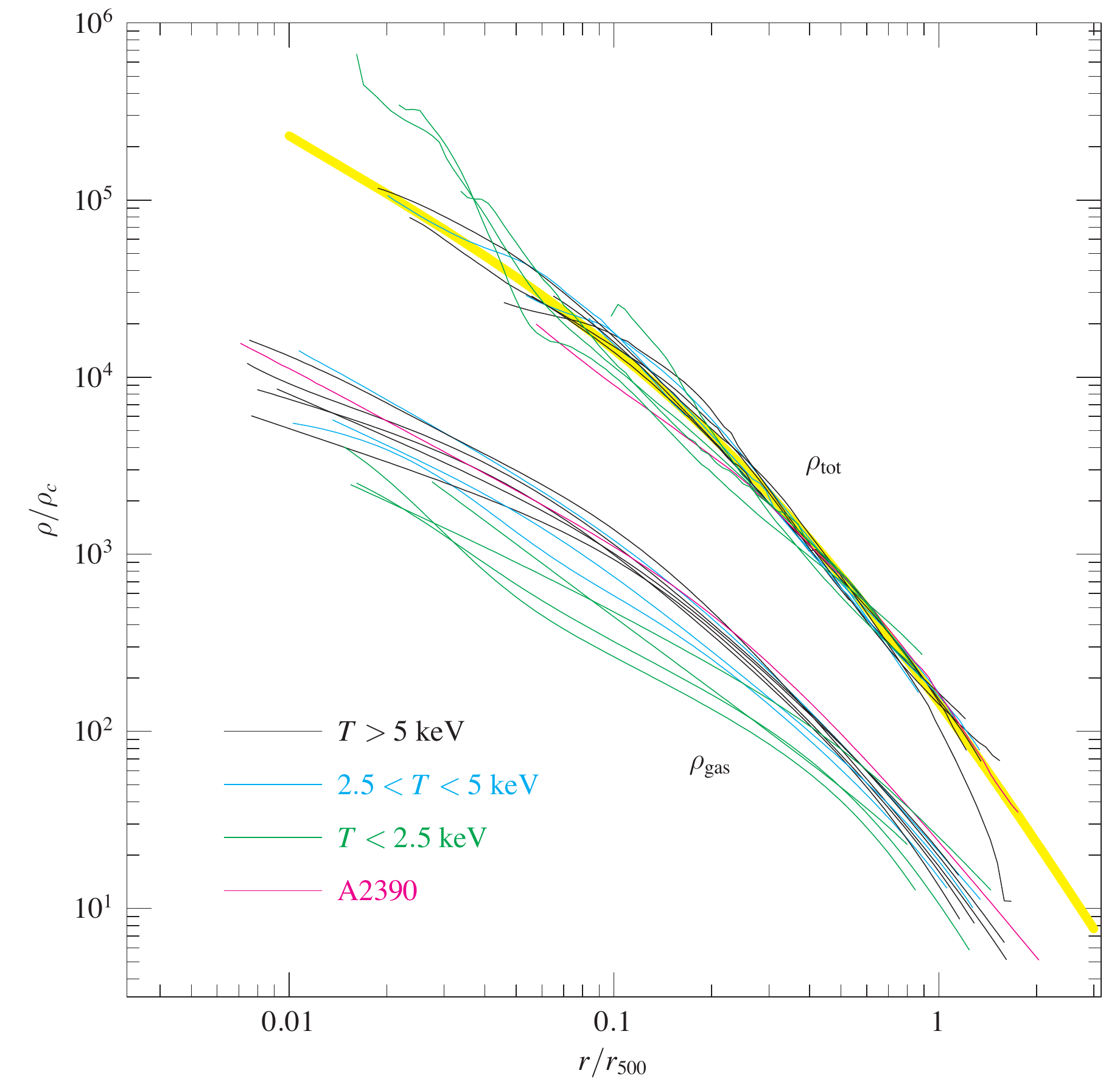}
\caption{{\it Left panel:} Rotational curve of the  galaxy NGC6503. I superimposed the optical image of corresponding galaxy with the   rotational curve, approximately to scale in radius. {\it Right panel:} Scaled cluster density profiles extracted from X-ray observations of different clusters, from Ref.~\cite{Vikhlinin:2005mp}.}
\label{RotCurves}
\end{center}
\end{figure}

Consider a test particle which is orbiting a body of mass $M$ at a distance
$r$. Within the frameworks of Newtonian dynamics the velocity of a particle is
given by
\begin{equation}
v_{\rm rot} =\sqrt{G\, M(r) \over r} \; .
\label{rot-velocity}
\end{equation} 
Outside of the body, the mass does not depend on distance, and the rotational
velocity should obey the Kepler law, $v_{\rm rot} \propto r^{-1/2}$.  Planets
of the Solar system obey this law. However, this is not the case for stars or
gas which are orbiting galaxies. Far away from the visible part of a galaxy,
rotational curves are still rising or remain flat. An example is shown in
Fig.~\ref{RotCurves},  left panel. An optical image of the NGC6503 galaxy is superimposed with
its rotational curve, approximately to correct scale. The contribution of
visible baryons in the form of stars and hot gas can be accounted for, and the
expected rotational curve can be constructed. The corresponding contributions
are shown in Fig.~\ref{RotCurves},  left panel. One can see that the data-points are far
above the contribution of visible matter. The contribution of missing dark
mass, which should be added to cope with data, is also shown and is indicated
as Dark Halo.  For the rotational velocity to remain flat, the mass in the
halo should grow with the radius as $M(r) \propto r$, i.e., the density of
dark matter in the halo should decrease as $\rho(r) \propto r^{-2}$.

\subsubsection{Dark matter density profiles.}

To interpret what is seen in the data, in particular, to interpret the results of  direct and indirect dark matter searches, 
and to plan for further strategy, it is important to know the expected 
phase-space structure of the dark halo and corresponding dark matter density profiles. 
For interacting  particles a thermal distribution over energies is eventually
established. However, in conventional cold dark matter models, particles
are non-interacting, except gravitationally. Binary gravitational interactions
are negligible for elementary particles, and resulting phase-space
distributions are not unique, even for a stationary equilibrium states, and even
if constraint to the flat rotational curves is enforced. Below I highlight several such distributions, 
which are often discussed in the literature and are used in applications. 

1. The simplest self-gravitating stationary solution which gives flat
rotational curves corresponds to an ``isothermal sphere'' with Maxwellian 
distribution of particles over velocities:
\begin{equation}
n(\vec{r},\vec{v}) = n(r)\; e^{-v^{2}/v^{2}_0}\; .
\label{n-isothermal}
\end{equation}
Solution of the equation of hydrostatic equilibrium can be approximated by the
density profile
\begin{equation}  \rho (r) = \frac{\rho_0}{(1 + x^{2})},~~~
{\rm where}~~   x \equiv r/r_c \; .
\label{rho-isothermal}
\end{equation}
It should be stressed that the distribution Eq.~(\ref{n-isothermal}), in
contrast to a distribution in real thermal equilibrium, depends on particle
velocities, not on their energies. Such distributions may arise in
time-dependent gravitational potential as a result of collisionless
relaxation.

2.  There exist several density profiles which are empirical fits to numerical
simulations, most often used is Navarro, Frenk \& White (NFW) profile \cite{Navarro:1997he}.
\begin{equation}
\rho (r) = \frac{\rho_0}{x\; (1 + x)^{2}}.
\label{eq:NFW}
\end{equation}

3. In the CDM model, the distribution of dark matter particles in the phase space during initial linear stage prior to structure formation corresponds to thin hypersurface, ${\bf v} = H {\bf r}$ (this is just Hubble law). Since during collisionless evolution the phase-space density conserves as a consequence of the Liouville theorem, then even  at the non-linear stage the distribution will still be a thin hypersurface. It can be deformed in a complicated way and wrapped around, but it cannot tear apart, intersect itself, puff up or dissolve. The resulting idealised phase-space distribution describing isolated dark halo has been studied  in  Ref.~\cite{Sikivie:1997nn}  for the case of spherical symmetry. Initial thin hypersurface gets  wrapped around indeed, forming large and ever increasing number of folds in the phase space   in the inner galaxy. Existence of such a folded structure is a topological statement.   Corresponding model is also called  "infall model". It  reproduces flat rotational curves, but there are several   interesting differences to other distributions, though.  Rotational curves of the infall model have several small ripples which appear near caustics in the dark matter distribution. (Those  are places in the phase-space where particles turn-around and have zero velocity).  It is important to note that caustics may be observable and their discovery in real data will prove that the missing mass is dark matter indeed, not a modification of gravity, even if dark matter particles will not be directly identified. The energy spectrum of dark matter particles at a fixed position also deviates from other  distributions. This may be important for  the direct dark matter searches. Also, the  infall  model gives the insight ~\cite{Sikivie:1997nn} on why the empirical fit, the NFW profile, has this particular form, Eq~(\ref{eq:NFW}).  Observationally, signature of the infall is seen e.g. in our local group of galaxies~\cite{Steigman:1998sb}, but at largest distances, outside of caustics. Caustics which are furthest away from the galaxy centre were resolved recently in the high-resolution N-body modelling of galaxy formation~\cite{Dolag:2012uq}. To understand how far the folded structure of the infall model continues into the inner halo in a galaxy like our own will require even larger simulations with better resolution.

\subsubsection{Dark Matter in Clusters of Galaxies}

Already in 1933, F. Zwicky \cite{Zwicky:1933gu} deduced the existence of dark
matter in the Coma cluster of galaxies. Nowadays, there are several ways to
estimate masses of clusters: based on the kinetic motion of member galaxies,
on X-ray data, and on gravitational lensing. These methods are different and
independent.  In the dynamical method, it is assumed that clusters are in
virial equilibrium, and the virialized mass is easily computed from the
velocity dispersion.  In X-ray imaging of hot intracluster gas, the mass estimates
are obtained assuming hydrostatic equilibrium. Mass estimates based on lensing
are free of any such assumptions.  All methods give results which are
consistent with each other, and tell that the mass of the luminous matter in
clusters is much smaller than the total mass. Recent review on basic properties of 
clusters and their role in modern astrophysics and cosmology can be found 
e.g. in~\cite{Vikhlinin:2014gfa}.

{\it Kinetic mass estimates.~}
Those are based on the virial theorem,~ $\langle E_{\rm pot} \rangle +
2\langle E_{\rm kin} \rangle = 0 $.  Here $ \langle E_{\rm kin} \rangle$
 is averaged kinetic energy of a constituents in the gravitationally
bound object (e.g. cluster of galaxies) and $ \langle E_{\rm pot} \rangle $ is their averaged
potential energy. Measuring the velocity dispersion of galaxies in the
clusters and its geometrical size gives an estimate of the total mass,
\begin{equation}
M \sim \frac{2 \langle r \rangle \langle v^{2} \rangle}{G} \; .
\label{virial-mass-estimate}
\end{equation}
The result can be expressed as mass-to-light ratio, $M/L$, using the Solar
value of this parameter. For the Coma cluster, which consists of about 1000
galaxies, Zwicky \cite{Zwicky:1933gu} has found
\begin{equation}
{M}/{L} \sim 300\, h\, \( {M_\odot}/{L_\odot}\; \) .
\end{equation}
Modern techniques end up with very much the same answer. 

{\it Mass estimates based on X-rays.~}
Mass is also traced in clusters of galaxies by the hot gas which is visible in
X-rays. Assume hot gas is in thermal equilibrium in a gravitational well
created by a cluster.  
Then, cluster mass profiles can be derived from the gas density and temperature 
as functions of the distance to a cluster centre. This independent method has its own 
advantages and disadvantages. With respect to galaxy dynamics (see above) or lensing
mass estimates (see below), this method has the advantage of being less sensitive to projection 
effects along the line of sight through the cluster. However, 
validity of the assumptions of ICM hydrostatic equilibrium and spherical symmetry of 
the cluster gravitational potential wells may depend on the evolutionary state of the cluster.

As an example, the radial density profiles  derived in~Ref.~\cite{Vikhlinin:2005mp} from the Chandra X-ray satellite  data are shown in Fig.~\ref{RotCurves},  right panel. We see that dark matter  density exceeds  gas density by an order of magnitude at any value of the radius. Dark matter density as a function of radius is well fitted by NFW profile, Eq.~(\ref{eq:NFW}), which is shown by thick yellow line. Total gas mass fractions  varies between 5 and 15 percent from cluster to cluster and systematically depends upon cluster mass. These values are somewhat lower than the Universal baryon fraction suggested by the CMB observations, but approaches it for the heaviest clusters.

\begin{figure}
\begin{center}
\parbox{0.5\textwidth}{\vspace{-5.3cm}
\includegraphics[width=0.52\textwidth]{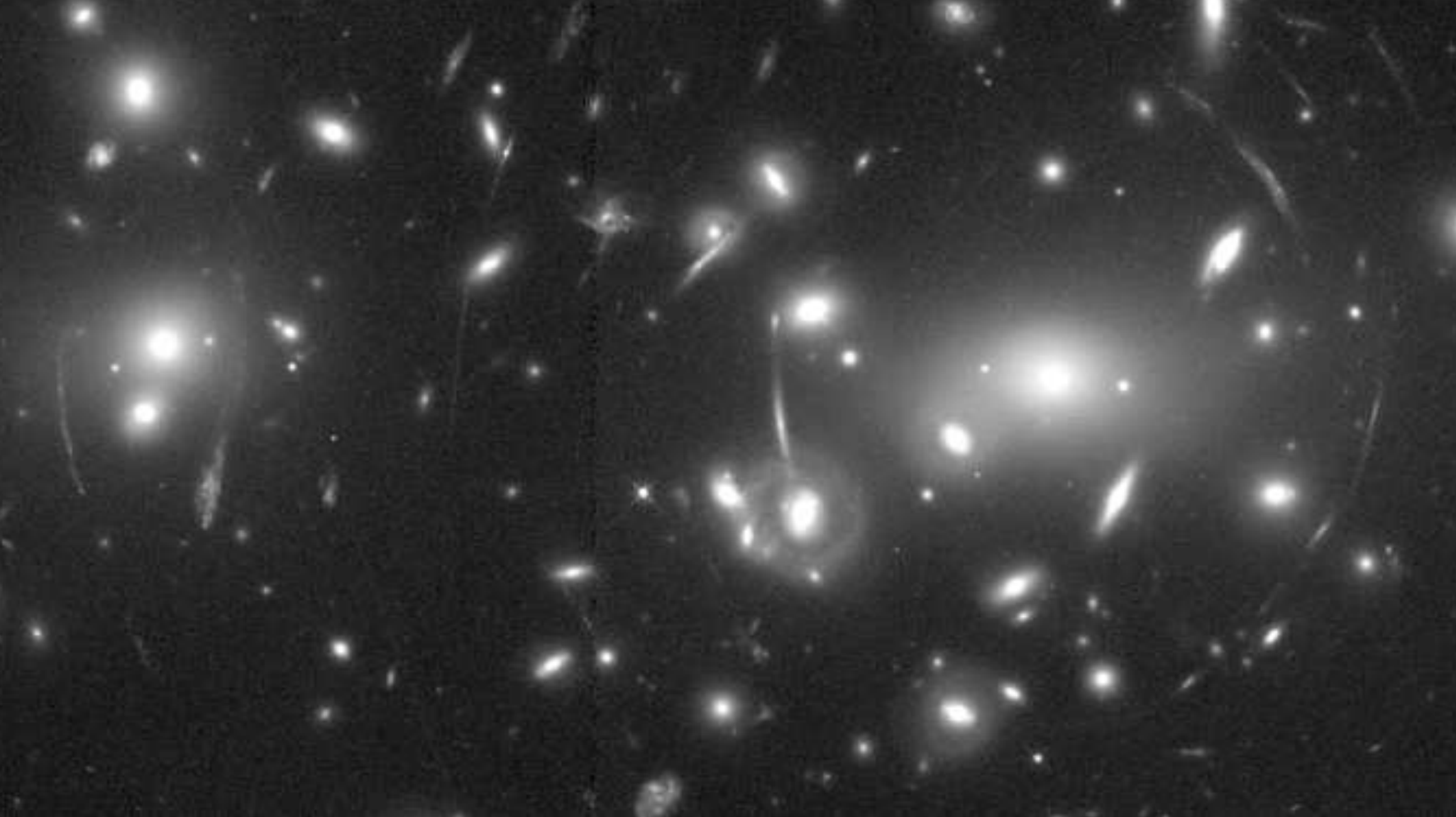}
}
\includegraphics[width=0.47\textwidth]{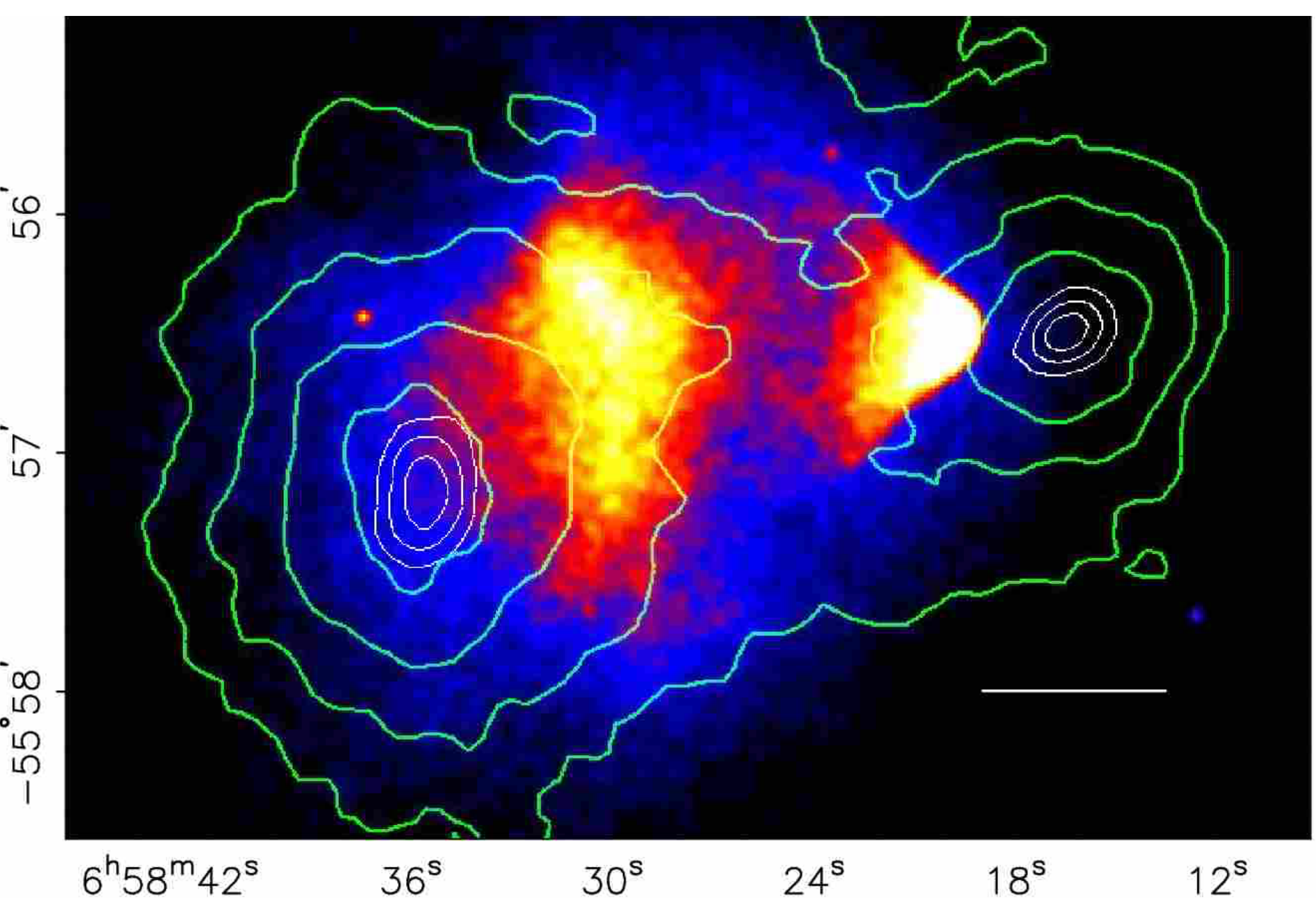}
\caption{{\it Left panel:} Image of the cluster Abell 2218 taken with the Hubble space telescope (see Ref.~\cite{Kneib:1995hh}). Spectacular arcs resulting from strong lensing of background galaxies by dark matter in the cluster are clearly seen. {\it Right panel:} Deep Chandra image of the Bullet cluster. The X-ray brightness of the gas component is coded  in yellow, red and blue colours. Distribution of the gravitating mass, obtained  from weak lensing  reconstruction,  is shown by green contours, from~Ref.~\cite{Clowe:2006eq}.}
\label{fig:GravLens}
\end{center}
\end{figure}

{\it Gravitational Lensing.~}
As photons travel from a background galaxy to the observer, their
trajectories are bent by mass distributions.  This effect of gravitational lensing allows direct mass measurement without any assumptions
about the dynamical state of the cluster. The method relies on the measurement
of the distortions that lensing induces on the images of the background
galaxies,
an example of such distortions is shown in  Fig.~\ref{fig:GravLens}, left panel.
A reconstruction of lens geometry provides a map of the mass distribution in
the deflector. For a review of the method see
e.g. Ref~\cite{Bartelmann:1999yn}.  The images of extended sources are
deformed by the gravitational field.  In some cases, the distortion is strong
enough to be recognized as arcs produced by a galaxy cluster serving as a lens,
see Fig.~\ref{fig:GravLens}, left panel.  For the cluster A 2218, shown in this
figure, Squires et al. \cite{Squires:1996ee} compared the mass profiles
derived from weak lensing data and from the X-ray emission.  The reconstructed mass
map qualitatively agrees with the optical and X-ray light distributions. A
mass-to-light ratio of $M/L = (440 \pm 80)h$ in solar units has been derived. 
The gas to total mass ratio was found to be $M_{\rm gas}/M_{\rm tot} = (0.04 \pm 0.02) \,
h^{-3/2}$. The radial mass profile agrees with the mass distribution
obtained from the X-ray analysis.  
For a recent study of mass density profiles of galaxy clusters derived
from  the gravitational lensing see e.g.~Ref.~ \cite{Okabe:2013efa}.  A sample of 50 galaxy
clusters at 0.15 < z < 0.3 has been studied.
Again, dark matter density as a function of radius is perfectly fitted by  the NFW profile, Eq.~(\ref{eq:NFW}), but 
"isothermal" profile is a bad fit.

{\it Dark matter or modification of gravity?~} In principle, the excess gravitational force, undoubtedly observed in galaxies and clusters of galaxies, could be not a manifestation of  the Dark Matter, but may have origin in some modification of Einstein gravity. Gravitational lensing studies of the Bullet Cluster 1E 0657-56  are claimed~\cite{Clowe:2006eq} to provide the best evidence to date for the existence of dark matter, as opposed do modifications of gravity. The Bullet Cluster consists of two colliding clusters of galaxies. Reconstructed distribution of the gas, stars and gravitating matters shown in Fig.~\ref{fig:GravLens}, right panel. The X-ray brightness of the hot gas  is coded  in yellow, red and blue colours. Distribution of the gravitating mass  is shown by green contours and was obtained  from weak lensing  reconstruction. It coincides with distribution of stars, but counts of stars gives small contribution to the overall mass balance. The hot gas of the two colliding components, seen in X-rays, represents most of the baryonic, i.e. ordinary, matter in the cluster pair. 
The hot gas in this collision was slowed down by a drag force. In contrast, the dark matter or stars were not slowed by the impact, because they do not interact strongly with itself or the gas except through gravity. This produced the separation of gravitating matter and gas seen in the data. If hot gas was the most massive component in the clusters, and dark matter would be absent, as proposed by alternative gravity theories, such a separation would not have been seen. Therefore, dark matter is required to explain what is seen here.

\subsubsection{Structure formation and DM}

By present time the structures in the Universe (i.e. galaxies and clusters) are
formed already, in other words perturbations in matter have entered non-linear regime, $\delta \rho /\rho \agt 1$.
However, the initial perturbations were small~~ $ \delta \rho /\rho \sim
10^{-5}$, as we know from measurements of temperature fluctuations in CMBR, see Section~\ref{sec:CMBR_tool}.  
Perturbations do not grow significantly in the radiation dominated epoch, they
can start growing only during matter domination and are growing then in proportion to the scale factor, 
$ \delta \rho /\rho \sim a = 1/z$.  Moreover, baryonic plasma is tightly coupled to radiation, therefore
perturbations in baryonic matter start to grow only after recombination. For
the same reason, initial perturbations in baryons at the time of recombination
are equal to fluctuations in CMBR. If baryons were to constitute the only matter
content, then perturbations in matter at present time would be equal to
\begin{equation}
\frac{\delta \rho}{\rho}|_{\rm today} = z_{\rm rec}\; \frac{\delta
\rho}{\rho}|_{\rm rec} \sim 10^{-2} \; , 
\end{equation} 
where $z_{\rm rec} \approx 1100$ is the redshift of recombination.  This apparent contradiction  is resolved by the dark matter. In our Universe
structure has had time to develop only because perturbations in non-baryonic
dark matter have started their growth prior to recombination. Baryonic matter
then ``catch up'' simply by falling into already existing gravitational wells.
This is one of the strongest and simplest arguments in favour of non-baryonic dark matter.

\subsection{Dark Matter: particle candidates}
\label{sec:DMcandidates}

Cosmology tells us that the Standard Model of particle physics is incomplete.
The model which will extend it should contain particles which would constitute non-baryonic dark matter. And there should exist some mechanism to produce it with correct abundance,  $\Omega_{\rm DM} \approx 0.27 $, see Table \ref{tbl:cosmoparameters}. Also, trusted and popular DM candidates appear naturally in the models whose  origin is unrelated to the dark matter problem.  There is no shortage of particle physics models which obey those requirements, with the huge range of DM paricle masses and very different production mechanisms. Some dark matter particle candidates are listed in Table~\ref{tbl:DMcandidates}. 
\begin{table}[h]
\begin{center}
\caption{Dark Matter particle candidates}
\label{tbl:DMcandidates}
\begin{tabular}{|l|l|l|}
\hline
candidate &  mass & some refs\\
\hline
{ ~Graviton} & ~$  10^{-21}$ eV & \cite{graviton}\\
{ ~Axion} & ~$ 10^{-5}$  eV & \cite{axion}\\
{ ~Sterile neutrino~~} & ~$  10$  keV &\cite{ster_neutrino}\\
{ ~Mirror matter} & ~$  1$  GeV &\cite{mirror}\\
{ ~WIMP} &  ~$  100$ GeV & \cite{WIMP}\\
{ ~WIMPZILLA} & ~$  10^{13}$  GeV~ &\cite{wimpzilla}\\
\hline
\end{tabular}
\end{center}
\end{table}
Given concrete model of particle physics, a theorist should first  calculate  the cosmological abundance of DM produced in the model in hands.  Below, in the subsection~\ref{sec:DM_production}, I give some examples of such calculations to highlight various mechanisms of DM production. Then, in the subsection~\ref{sec:DM_searches}, I briefly describe vast topic of direct and indirect searches for most popular DM candidates, with corresponding derived constraints.

\subsection{Production mechanisms}
\label{sec:DM_production}

Depending upon production mechanism, the resulting dark matter can appear as 'cold', 'warm' or 'hot'.
Loosely speaking, velocities of cold dark matter are so small that they are not  influencing the large scale structure formation at all. Velocities of  hot dark matter particles are too big. Their kinetic energy does not allow particles to clump galaxy halos  and may smear out even  clusters of galaxies. Such DM \ is ruled out. Warm dark matter is the intermediate case. It may wash out structure at smallest observable scales of dwarf galaxies  but does not influence formation of big haloes like our Milky Way. Cold dark matter models have some problems explaining  observations at small scales, the warm dark matter models have some advantages here, see below.

Further, dark matter particle candidates can be divided into several classes according to a mechanism of their production in the early universe. We start with popular class of DM candidates referred to as "thermal relics".

\subsubsection{Cosmological abundance of thermal relics}
\label{sec:ThRel}

By definition, a thermal relic is assumed to be in  thermodynamic equilibrium  at early times. At some point in the evolution particles go out of equilibrium and after that their number in a comoving volume remains constant. The process is called  "freeze-out".  For thermal relics it is just the value of particle mass which determines if it will be hot, warm or cold.
To see this, let us define the  {\it free streaming length} for a given DM particle species  with mass $M_X$ as a horizon size at a temperature when particles are still relativistic, i.e. at $T \sim M_X$. Clearly, structure will be washed out at all scales smaller than this. Later on, particles are non-relativistic and cannot move much farther away. Structure is preserved at larger scales. Horizon size at $T \sim M_X$ expanded to present epoch is given by
$$ L_{fs} \sim \frac{M_{\rm Pl}}{T_0 M_X}.$$
For ~$ M_X \sim 1~{\rm eV}$ this gives  $L_{fs} \sim 100~ {\rm Mpc}$. 
Clearly, models with such a big free streaming length are ruled out. On the other hand,
for ~$ M_X \sim 1~{\rm keV}$ we find  $L_{fs} \sim 0.1~ {\rm Mpc}$. This corresponds to the size of a dwarf galaxy. 
Therefore, this gives the  lower bound for the warm DM particle mass:
$$M_X > 1~{\rm keV}.$$

For thermal relics the resulting dark matter will be definitely cold if freeze-out occurs when particles are non-relativistic, i.e. at temperatures smaller than particle mass, $T <M_X$.
WIMP, and in particular neutralino, appearing in supersymmetric models, belong to this class of dark matter.

\begin{figure}
\begin{center}
\includegraphics[width=0.5\textwidth]{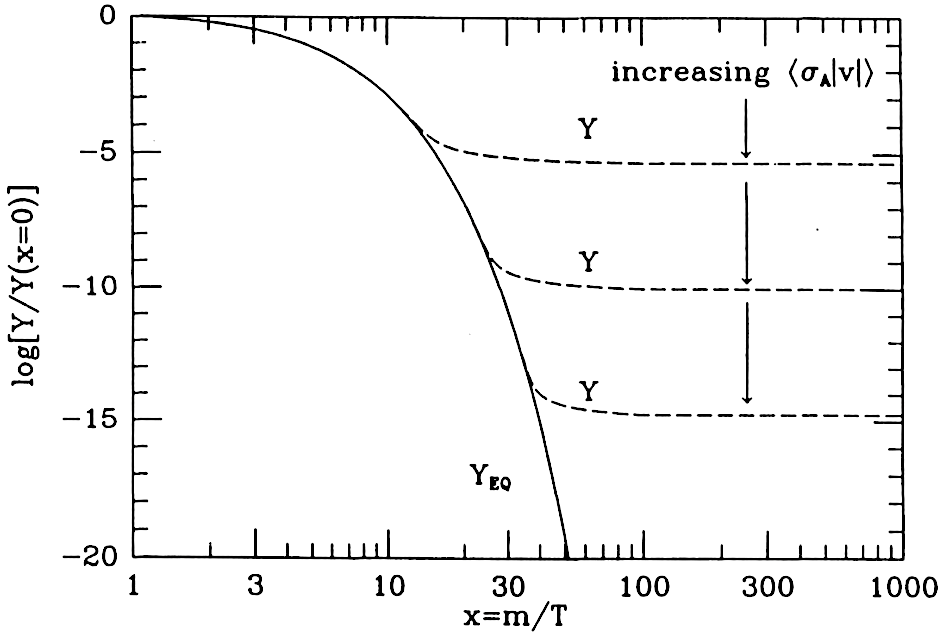}
\caption{A schematic view of comoving number density of a stable species as they evolve
through the process of thermal freeze-out.
}
\label{fig:freeze}
\end{center}
\end{figure}

A given particle species will track the equilibrium abundance as long as reactions which keep them in chemical equilibrium can proceed rapidly enough. Here, "rapidly enough" means that the mean free time between interactions is smaller than the age of the universe, $\tau < t_u$. This condition can  also be written as $ n\sigma v > H$. In thermal equilibrium, the number density of non-relativistic particles is given by Eq.~(\ref{eq:NNonRel}).
In this regime the number density decreases exponentially fast  with decreasing temperature $T$. When the rate of reactions becomes lower than the expansion rate, the particles can no longer track the equilibrium value and thereafter particle concentrations in a comoving volume remain constant.  Clearly, the more strongly interacting particles are, the longer they stay in equilibrium, and the smaller their freeze-out abundance will be, see Fig.~\ref{fig:freeze}. Here we defined particle abundance as the ratio of particle number to the entropy density, $Y \equiv n/s$.

Freeze-out concentration $n$ is determined by the condition
$ n_f\sigma v \approx H$, or (neglecting numerical factors)
$$ n_f \approx \frac{H}{\langle \sigma v \rangle} \approx \frac{T_f^2}{M_{Pl}\langle \sigma v \rangle}.$$
After freeze-out the ratio of $n$ and entropy density $s$ remains constant. In particular, present density is given by $n_0= n_f s_0/s_f$. Therefore
$$ \Omega_{DM}^{~} \equiv \frac{m n_0}{\rho_c} = 
\frac{m n_f}{s_f}\frac{s_0}{\rho_c}
\sim \frac{m}{T_f}\frac{1}{\langle \sigma v \rangle}\frac{T_0^3}{M_{Pl}\rho_c}.$$
Freeze-out temperature $T_f$ cannot go much below particle mass m, see Fig.~\ref{fig:freeze}. One gets $x_f \equiv m/T_f = 20 - 30$ for all practically interesting  values of annihilation cross-section.
Restoring now  all numerical factors in the above  estimate we obtain
\begin{equation} 
\Omega_{DM}^{~} = \frac{16\pi^2}{3}\,\sqrt{\frac{\pi}{45}}\,
\frac{x_f^{~}g_0^{~}}{\sqrt{g_*(T_f)}}\,
\frac{T_0^3}{M_{Pl}^3H_0^2}\;\frac{1}{\langle \sigma v \rangle}.
\label{eq:Om_vs_crs}
\end{equation}
For the $s$ - wave annihilation $\sigma = \sigma_0^{~}/v$ and we have numerically
$$ \Omega_{DM}^{~} \approx 0.2\, \frac{\rm pb}{\sigma_0^{~}}.$$
Note that a picobarn crossections are in the ballpark of the electroweak scale,
${\rm pb} \approx \alpha^2 /({100 \rm \; GeV})^{2}$. That is why the weakly interacting massive particles, appearing e.g. in supersymmetric extensions of the Standard Model are considered to be natural candidates for the dark matter.
Another useful parametrisation of this result is given by
\begin{equation}
\Omega_{DM}^{~} \approx 0.2\;\, \frac{3\cdot 10^{-26}\; {\rm cm}^3\;{\rm s}^{-1}}
{\langle \sigma v \rangle}.
\label{eq:Om_DM_f}
\end{equation}
This expression is used for the discussion of dark matter direct and indirect search results and strategies.

\subsubsection{Cosmological abundance of ultra-light bosons}

Dark matter particles can be very light and still very cold if  they did not originated from the thermal bath. Of course this holds for bosons only, since the phase-space restrictions will not allow light fermions to saturate required energy density in galaxy halos.  Corresponding constraint on fermions is called  Tremaine-Gunn limit \cite{Tremaine:1979we} and  reads $M_F \agt $ 1 keV. 

To illustrate the general idea, let us consider a scalar
field with potential $V(\phi ) = m^2\phi^2/2$. The field equations  for the
Fourier modes with a momentum $k$ in an expanding Universe are
\begin{equation}
\ddot{\phi}_k + 3H\dot{\phi}_k + (k^2 + m^2)\phi_k = 0\; .
\label{eq-motion-scalar}
\end{equation} 
Since the term $\propto H$ can be understood as a friction, amplitude of 
modes with $9H^2 \gg(k^2 + m^2) $ (almost) does not change with time. Then, the
oscillations of modes with a given $k$ commence when $H$ becomes sufficiently
small, $9H^2 \ll (k^2 + m^2)$. Oscillating modes behave like particles, and
their amplitude decreases with expansion. Since modes with the largest $k$
start oscillations first, they will have the smallest amplitude and the field
becomes homogeneous on a current horizon scale. This holds while mass term is 
 unimportant, i.e. till $3H > m$. Modes with all $k  < 3H$ will start oscillations simultaneously
 when $3H \approx m$, and will behave like cold dark matter
since then.

Resulting abundance of dark matter will depend upon initial amplitude of modes with
$k  < 3H$. Why the initial amplitude of such modes is non-zero in the first place? Such fields
are generated during inflation if $m$ is smaller than the value of the Hubble constant 
during inflation, see Eq.~(\ref{ModEq-massles-conf}). In this way e.g. massive gravitons 
are created as a dark matter, see Ref.~\cite{graviton}.

Situation in the case of axions is even easier to understand. Potential for the axion field $a$ has the following form
$$
V(a) = m_a^2 f_a^2\, \left( 1-\cos (a/f_a) \right).
$$
Axion mass is temperature dependent, $m_a = m_a(T)$ and at $T \gg 1$ GeV it is zero. Therefore, at this temperatures $V(a) = 0$ and the axion field takes arbitrary values in the range $0 <   a/f_a < 2\pi$. Field oscillations start with amplitude $a\sim f_a$ at $T \approx 1$ GeV when  $3H(T) = m_a(T)$. Correct axion abundance is obtained for $10^{-5} \lsim m_a \lsim 10^{-3}$. 
Note that the field will be homogeneous on the horizon scale at $T >1$ GeV, but may be inhomogeneous on larger scales. This may lead to formation of dense clumps, ``axion miniclusters'' of the mass $M
\sim 10^{-12}\,M_\odot$ \cite{Kolb:1993zz}.

\subsubsection{Cosmological abundance of superheavy particles}

Superheavy particles can be created purely gravitationally.
As we have seen in Section~\ref{sec:creation},  generically, a quantum field cannot be kept
in a vacuum in the expanding universe. This can be understood on the example of
a scalar field, Eq. (\ref{eq-motion-scalar}). In conformal time $\eta$,
Eq.~(\ref{eq:time-conformal}), and for rescaled field, $u_k \equiv \phi_k\, a$,
the mode equations take form of an oscillator equation
\begin{equation}
\ddot{u}_k ~+~ [k^2 + m^2_{\rm eff} (\eta)]\; u_k = 0 \, ,
\label{eq-motion-scalar-conformal}
\end{equation}
with time-dependent mass
\begin{equation}
m^2_{\rm eff} (\eta) =  a^2 m^2 - \frac{\ddot{a}}{a}(1-6\xi) .
\label{eq-motion-scalar-conformal}
\end{equation}
This is one particular case of the general situation described by Eq.~(\ref{ModEq}).
The constant $\xi$ describes the coupling to the scalar curvature, the
corresponding term in the Lagrangian is $\xi R \phi$. The value $\xi=0$
corresponds to minimal coupling (Eq.~(\ref{eq-motion-scalar}) was written for
this case), while $\xi=1/6$ is the case of conformal coupling. Equations for
massless, conformally coupled quanta are reduced to the equation of motion in
Minkowski space-time. Particle creation does not occur in this case. For
massive particles, conformal invariance is broken and particles are created
regardless of the value of $\xi$. Let us consider the case of $\xi=1/6$.  It is the particle
mass which couples the system to the background expansion and serves as the
source of particle creation in this case. Therefore, we expect that the number
of created particles in comoving volume is $\propto m^3$ and the effect is
strongest for the heaviest particles. In inflationary model~(\ref{chaotic}) the abundance of created particles, $\Omega_{\rm SH} $,  will match observations if $m \sim 10^{13}$ GeV  \cite{wimpzilla}, precise value of required superheavy particle 
mass depends upon  reheating temperature and the value of $\xi$. Therefore, a dark matter can be created in the early Universe even if it has no couplings at all, the only condition reads:  be superheavy.

\subsubsection{Cosmological abundance of sterile neutrino}
\label{sec:RHnu_asDM}

Active neutrino are massive, this fact signifies a new physics beyond the Standard Model. Other fermions have masses because they exist as left handed and right handed states with coupling to the Higgs field H. However, active neutrinos are left-handed. Therefore, a natural way to generate masses for the neutrino would be to consider them at the same footing as other fermions and to add right handed neutrinos, ${N_j}$, to the Standard Model Lagrangian, 

\begin{equation}
  {\cal L}
  = {\cal L}_{\rm{SM}} + i\bar{N_j}  \partial_\mu \gamma^\mu N_j
  - \left[\lambda_{ji} (\bar{L_i}H) N_j 
  + \frac{M_{j}}{2} \; \bar N_j^cN_j + \rm{h.c.} \right] .
 \label{eq:NuMSM}
\end{equation}
Flavour indexes $j$ may run from one to three, but not necessarily. In what follows I omit explicit writing of indexes. In the first term in square brackets $L_i$ stands for a doublet of left-handed leptons. This term generates Dirac masses for the neutrino, $m_D = \lambda  \langle H \rangle$. In general, right handed neutrino may have  Majorana masses, $M$, as well. Such term is forbidden for other fermions since their right-handed components have charges, but right handed neutrino are neutral. 
 
Right-handed components are also called {\it sterile neutrino} since they do not interact directly with particles of the Standard Model. However, they are not really sterile since interact with other particles via mixing. Indeed,
to get neutrino mass eigenstates we have to diagonalise mass matrix in square brackets of Lagrangian (\ref{eq:NuMSM}). This gives mixing of active and sterile neutrino
\begin{equation}
\theta = \frac{m_D}{M}.
\label{eq:NuMixing}
\end{equation}
Therefore, sterile neutrino interaction matrix elements are the same as for the active neutrino except they are multiplied by $\theta$. If ${m_D} \ll {M}$ the masses of heavy states nearly coincide with $M$ and the lightest among  sterile neutrinos is a good candidate for dark matter if its mass $M \agt 1$ keV. (But not heavier than 50 keV, otherwise its decays to $\gamma$ will contradict observed X-ray astrophysical backgrounds, see Section~\ref{sec:DM_searches}.)

Sterile neutrino can be produced in the early Universe  directly in the inflation decays \cite{Shaposhnikov:2006xi}, or
via mixing, Eq. (\ref{eq:NuMixing}), with active neutrino \cite{Dodelson:1993je}. Production rate of sterile neutrinos in the latter case can be obtained multiplying production rate for the active neutrinos in primordial plasma by mixing angle squared
$$\Gamma  \approx \theta^2\sigma^{~}_W n \sim \theta^2G_F^2 T^2\cdot T^3.$$
Multiplying this rate by time, $t \sim H^{-1} \sim {M^{~}_{\rm Pl}}/{T^2}$ we obtain  number density of sterile neutrinos produced
$$\frac{n_s^{~}}{n_\gamma} \sim \theta^2\, G_F^2 T^3 M^{~}_{\rm Pl}.$$
To close this estimate we note that active-sterile neutrino mixing is temperature dependent \cite{Dolgov:2000ew}
$$\theta \rightarrow \theta_M = \frac{\theta}{1+2.4(T/200~{\rm MeV})^6({\rm keV}/M_1)^2},$$
which gives for the production temperature of sterile neutrino
$$T \sim 130 \left(\frac{M}{1~{\rm keV}}\right)^{1/3}~ {\rm MeV},$$
and resulting abundance \cite{Dodelson:1993je}
\begin{equation}
\Omega_{s} \sim \Omega_{m} \frac{\sin^2(2\theta)}{10^{-7}} \left(\frac{M}{1~\rm keV}\right)^2,
\label{eq:NuAbondance}
\end{equation}
where $\Omega_{m}$ is observed  dark matter abundance. 
Proper calculation involves solution of Boltzmann equations. Details and the list of references can be found in the recent review \cite{Adhikari:2016bei}. Quoted result, Eq.~(\ref{eq:NuAbondance}), corresponds to zero lepton asymmetry. With maximum asymmetry the required $\theta$ can be two orders of magnitude smaller  \cite{Laine:2008pg} at the same mass of sterile neutrino, see Fig.~\ref{fig:nu-bounds}, right panel.

\subsection{Dark matter searches and constraints}
\label{sec:DM_searches}

Dark matter particles, in majority of suggested models,  can be discovered in direct  dedicated searches in  laboratories.
Dark matter can also leave trace and be identified  in indirect searches, e.g. analysing data on cosmic ray, X-ray, gamma-ray and neutrino telescopes.

\begin{figure}
\begin{center}
\includegraphics[width=0.46\textwidth]{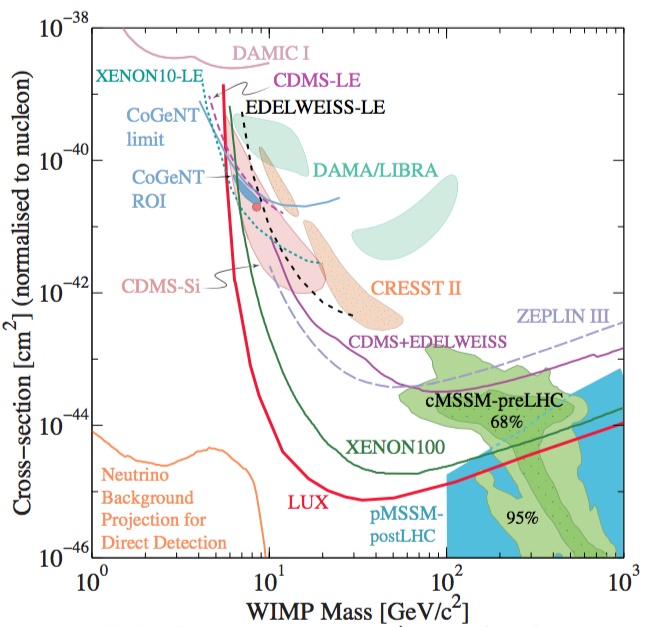}
\includegraphics[width=0.53\textwidth]{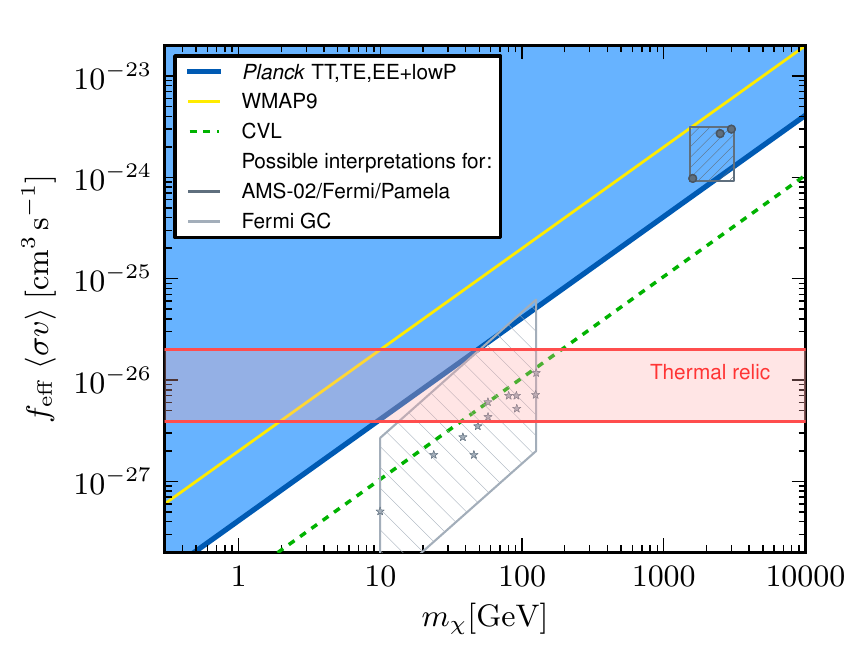}
\caption{Left panel: Constraints on neutralino from different underground  experiments are shown by correspondingly marked colored lines. Similarly, claimed hints of detection are represented by shaded areas.  Remaining expectation for supersymmetric models are shown by shaded area in the lower right corner, marked as MSSM. Right panel: 
Constraints on neutralino self-annihilation cross-section from CMBR. The blue area shows the parameter space excluded by the Planck data. The yellow line indicates the constraint using WMAP9 data. The dark grey circles and light grey stars correspond to various claims of indirect detection as cosmic or $\gamma$-ray excesses. The horizontal red band corresponds to correct neutrino abundance, Eq.~(\ref{eq:Om_DM_f}).  From Ref.~\cite{Ade:2015xua}.}
\label{fig:wimp-bounds}
\end{center}
\end{figure}

1. {\it Neutralino. } WIMP particles have tiny but phenomenologically important elastic cross-section with usual baryonic matter.  For WIMPs heavier than nuclei, $m_X \gg m_N$ and a typical velocities in the Galaxy halo, $v \sim 300$ km/s, typical recoil energy is $E_R \sim m_N v^2 \sim 1 \;-\; 100$ keV. The recoil can be measured studying  ionization, scintillation, heat or sound waves it creates in a detector.
Different experiments use different techniques, or their combinations.  But, of course, it should be done deep underground, in low background laboratories. Current observational bounds on the scattering cross-section exclude a lot of the WIMP parameter space of  MSSM but do not test the bulk of it, see Fig.~\ref{fig:wimp-bounds}, left panel.  Intriguingly,  crystal-based experiments CDMS Si, CoGeNT ROI, CRESST II and DAMA/LIBRA claim some hints of a positive dark matter signal. These claims  are mutually exclusive, and cluster in the mass region of tens of GeV and at cross sections between $10^{−42}$ and $10^{−39}$ cm$^2$, see Fig.~\ref{fig:wimp-bounds}, left panel. However, the noble-gas experiments ZEPLIN,  XENON and  most recent LUX, exclude this parameter region. Remaining expectation for supersymmetric models, after all constraint are taken into account, including LHC results, are shown by shaded area in the lower right corner, marked as MSSM, in the same figure. The uncertainty  for the expected signal arises because   the scattering cross-section is not directly related to the annihilation cross-section. 

However, that same self-annihilation that plays a central role in the freeze-out, see Section~\ref{sec:ThRel},
 leads also to the dark matter annihilation in the Galaxy halo. It can give rise to a significant flux of $\gamma$-rays, neutrinos, and even antimatter such as antiprotons and positrons, especially from regions with large dark-matter density . This creates prospective signal for the indirect WIMP detection. It is searched for, as an excess over conventional astrophysicsl backgrounds, by the orbital cosmic ray observatories, ground based atmospheric Cherenkov and neutrino telescopes. Though annihilation cross-section for indirect searches is fixed, some uncertainty arises here because of certain uncertainty in dark matter density profiles. 
 
 Dark-matter annihilation with a non-vanishing branching ratio into the electromagnetic channel leads also to distortions of the CMB which has been probed with WMAP and Planck data~\cite{Ade:2015xua}.  WIMPs lighter than 10 GeV originating in  thermal freeze-out scenario are excluded by these observations and the advantage of CMB-based limits lies in the absence of astrophysical uncertainties, see Fig.~\ref{fig:wimp-bounds}, right panel.  Dark matter annihilation interpretation of the cosmic ray excess detected by AMS, Fermi and Pamela satellites (shown by  dark grey circles) are also excluded now by Planck.  However,  the interpretation of the $\gamma$-ray excess from the Galactic centre measured  by Fermi (corresponding parameter regions are indicated by light grey stars) is still viable. Intriguingly, it intersects with the horizontal red band which corresponds to the correct neutralino abundance as thermal relic. 
 
Dark matter particles escape direct detection at colliders such as the Large Hadron Collider (LHC) at CERN, however, they would produce a characteristic signal of missing energy. Arising  constraints on WIMP-nucleon cross-section are model dependent, but are powerful and competitive with direct searches in underground labs, especially in the region of low masses.
 Recent detailed review on direct, indirect and collider WIMP searches can be found e.g. in Ref.~\cite{Klasen:2015uma}.

\begin{figure}
\begin{center}
\includegraphics[width=0.48\textwidth]{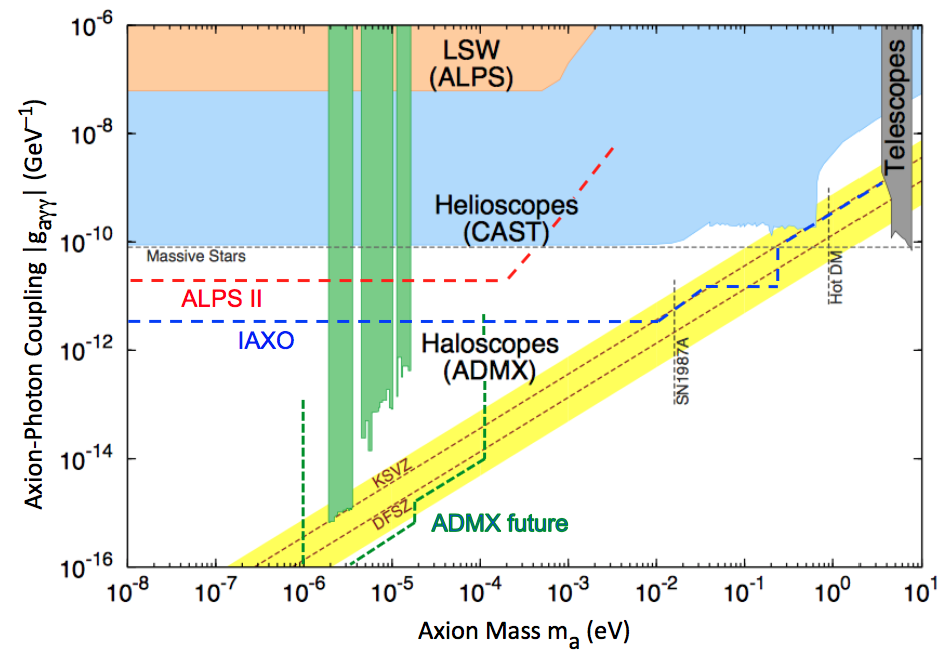}
\includegraphics[width=0.49\textwidth]{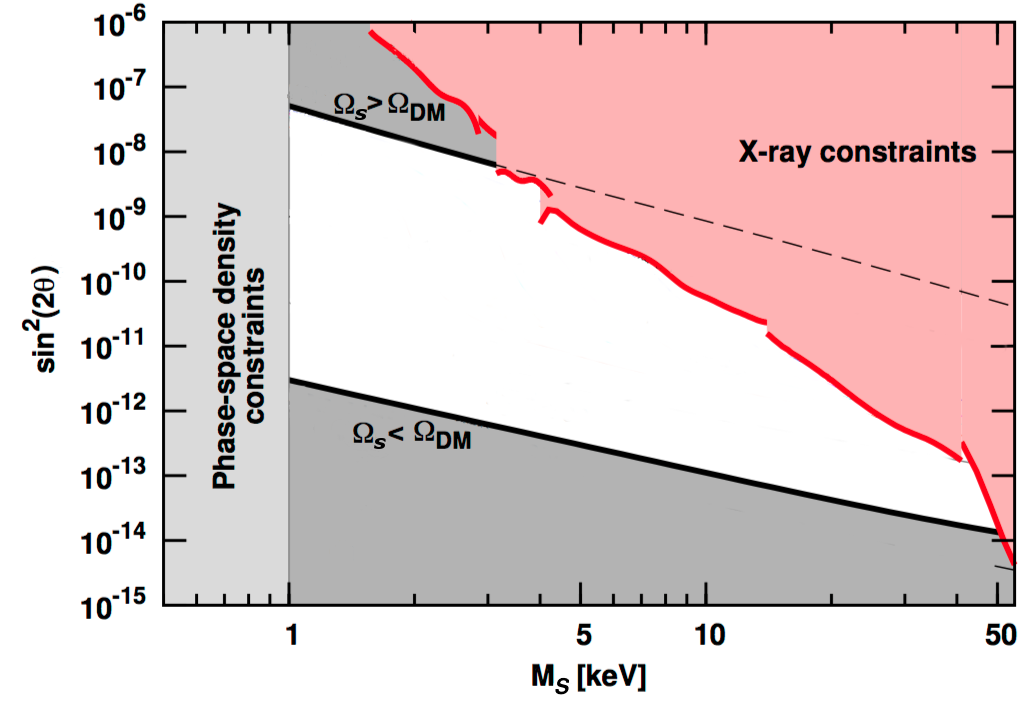}
\caption{Left panel:  Parameter space for the axion and axion-like dark matter models. Yellow band corresponds to the correct cosmological abundance of QCD axions if they make all of  dark matter. Regions excluded by ADMX and CAST are shown by green and blue shaded areas. Dashed lines show prospective limits of planned experiments. From Ref.~\cite{Graham:2015ouw}. Right panel:  Unshaded white region represents allowed parameter space for DM sterile neutrinos. The upper and  lower thick black lines correspond to correct abundances for zero and maximal lepton asymmetry. Red  region in the upper right corner is forbidden by the  X-ray constraints. The region below 1 keV is ruled out by the phase-space density arguments.  Adapted from Ref.~\cite{Boyarsky:2009ix}.}
\label{fig:nu-bounds}
\end{center}
\end{figure}

2. {\it Axions. } 
Axion interactions with photons and fermions can be parametrised as 
\begin{equation}
 L_{\rm int}=- \frac{1}{4}g_{a\gamma} \, a F_{\mu \nu} \tilde F^{\mu \nu} - \sum_{\rm fermions}g_{ai} \, a \overline \psi_i \gamma_5 \psi_i \,,
\label{eq:axion_coupling}
\end{equation}
where
$$
g_{a\gamma} \equiv \frac{\alpha}{2\pi} \frac{C_{a\gamma}}{f_a}, ~~~~~~ g_{ai} \equiv \frac{m_i}{f_a }C_{ai},
$$
and $ C_{a\gamma}$,  $ C_{ai} $ are model dependent parameters (in simple models of order unity). Direct axion searches in the laboratory are based on interactions with $\gamma$. Namely, axions constituting the Milky Way dark matter halo would resonantly convert into a monochromatic microwave signal in a high-Q microwave cavity permeated by a strong magnetic field \cite{Sikivie:1983ip}. Such axion search experiments (ADMX is the most recent one) are called {\it haloscopes}.
Similarly, axions or axion-like particles, emitted by the Sun will be converted in the strong magnetic field into X-ray photons. Axion experiments which search for this signal (CAST is the most recent one) are called {\it helioscopes}. Constraints  obtained by ADMX and CAST experiments are shown in Fig.~\ref{fig:nu-bounds}, left panel. 

Interaction Lagrangian Eq.~(\ref{eq:axion_coupling}) leads also to a new observable astrophysical phenomena, which may lead to indirect axon detection and give constraints on axion parameters. Extra energy losses by starts is one of those effects. Corresponding constraints practically coincide with the bound obtained by CAST. It is shown by the dotted line marked by the label "Massive Stars" in the Fig.~\ref{fig:nu-bounds}, left panel.  No accident, along the same line we can find models capable to explain several claims hinting for the axion effects in the astrophysical data, for a review see e.g. Ref.~\cite{Ringwald:2015lqa}.
Recent detailed review on direct and indirect axion searches can be found also in Refs.~\cite{Klasen:2015uma,Graham:2015ouw}.

3. {\it Sterile neutrino. } 
Recent detailed review on direct and indirect sterile neutrino searches can be found in Ref.~\cite{Adhikari:2016bei}.
In every process where active neutrino appears, sterile neutrino can appear as well, again via mixing, Eq. (\ref{eq:NuMixing}). This opens the way for a laboratory sterile neutrino searches. For example, in the keV mass range, appearance of sterile neutrinos changes kinematics and the spectrum of nuclear decays. Most recent searches of sterile neutrinos in tritium  $\beta$-decay has started in Troitsk  \cite{Abdurashitov:2015jha}, and will be continued at KATRIN experiment~\cite{Adhikari:2016bei}.

Also, at one loop level sterile neutrino are decaying into active neutrino and photon. Loop diagrams for this process are the same as for the electromagnetic form factors of a massive neutrino in the Standard Model with one external neutrino leg being connected to sterile neutrino via mixing (\ref{eq:NuMixing}). The decay width can be easily obtained using e.g. results of Ref.~  \cite{Pal:1981rm} and is given by 
\begin{equation}
\Gamma_{N\rightarrow \gamma \nu_a} = \frac{9\,\alpha \, G_F^2}{256\cdot 4\pi^4}\sin^2 2\theta \, m_s^5 =  5.5\times 10^{-22}\,\theta^2 \left[\frac{m_s}{1~\rm keV}\right] ^5~ {\rm s}^{-1} .
 \label{eq:NuWidth}
\end{equation}
Because of that, sterile neutrino dark matter is not completely dark. It can be detected by searching for an unidentified X-ray line, which would appear at a frequency $\omega = m_s/2$. Intensity of this line should follow dark matter density profiles. Dwarf satellite galaxies are a good places to search for such a signal because they are dark matter dominated and usual  astrophysical X-ray background is small there \cite{Boyarsky:2006fg}.

To conclude this section. A large number of various clams exists in the literature with a hints of indirect dark matter signal for all of the candidates described above: nutralino, axion-like particles and sterile neutrino. Do describe these hints in detail would require separate volume, interested reader can consult recent reviews \cite{Klasen:2015uma,Ringwald:2015lqa,Adhikari:2016bei}. As usual, hints appear at a boundary of allowed parameter space where observational capabilities are stretched. Moreover, indirect dark matter signal can be confused with  conventional  astrophysical  backgrounds or effects. Clearly, these claims are mutually exclusive and it is not possible for all of them to be precursors of the true signal, since dark matter is either neutrino, or axions, or sterile neutrino, or something else. On the other hand, one of those may turn out to be true and it is not excluded that we see already the tip of a real iceberg.

\section{Conclusions}

As we have seen, cosmology and astrophysics gave us solid evidence that the Standard Model of particle physics is incomplete. We have  to extend it to explain neutrino masses, baryogenesis,  and dark matter. Dark energy can be explained by the Einstein's $\Lambda$-term, but we do not know why it exists, and there seems to be too many coincidences between numerical values of cosmological parameters. On the other hand, a form of  dark energy explains the Universe origin within inflationary paradigm, which increasingly finds support  in cosmological data.

Cosmology just recently became a precision science but is full of surprises already, helping to build true model of microphysics.  It is up to high energy physicists  to find out what this new physics is. With advances of this program we, in turn, will have better understanding of the Universe origin, of its evolution, of its current  state, and of its future fate.

\section*{Acknowledgements}

I would like to use the opportunity to thank CERN and Dubna  for the organization and support of this annual summer school series for the young high-energy physicists.

\appendix

\section{Gravitational creation of metric perturbations}

As an important and simple example, let us consider quantum fluctuations of a
real scalar field, which we denote as $\varphi$. It is appropriate to rescale
the field values by the scale factor, $\varphi \equiv \phi/a(\eta)$. This
brings the equations of motion for the field $\phi$ into a simple form of
Eq.~(\ref{ModEq}).  As usual, we decompose $\phi$ over creation and
annihilation operators $b_{\k}$ and  $b_{\k}^{\dagger}$
\begin{equation}
\phi(\x,\eta) = 
\int \frac{d^3k}{(2\pi)^{3/2}}\; \left [ u_k(\eta)\, b_{\k}\, \e^{i\k\x} + 
  u^{*}_k(\eta)\, b_{\k}^{\dagger}\, \e^{-i\k\x}   \right] \; .
\label{field-fourier-decomp} 
\end{equation}
Mode functions $u_k$ satisfy Eq.~(\ref{ModEq}). In what follows we will assume
that $\varphi$ is the inflaton field of the ``chaotic'' inflationary model,
Eq.~(\ref{chaotic}). During inflation $H\gg m$ and $H \approx {\rm const}$.
So, to start with, we can assume that $\varphi$ is a massless field on the
constant deSitter background. (The massive case can be treated similarly, but
analytical expressions are somewhat more complicated and do not change the
result in a significant way. Corrections due to change of $H$ can also be
taken into account, and we do that later for the purpose of comparison with
observations.)  With a constant Hubble parameter during inflation the solution
of Friedmann equations in conformal time is
\begin{equation}
a(\eta ) = - \frac{1}{H\eta} 
\label{a-desitter-confomal} 
\end{equation} 
and the equation for mode functions of a massless, conformally coupled to
gravity $(\xi =0)$, scalar field takes the form 
\begin{equation} 
\ddot{u}_k + k^2 {u}_k -\frac{2}{\eta^2}\; {u}_k = 0 \; .
\label{ModEq-massles-conf}
\end{equation}
Solutions which start as vacuum fluctuations in the past ($\eta \rightarrow
-\infty$) are given by
\begin{equation}
{u}_k = \frac{\e^{\pm ik\eta}}{\sqrt{2k}}\left( 1 \pm
\frac{i}{k\eta}\right) \; .
\label{ModFunc-massles-solution}
\end{equation}
Indeed, at $\eta \rightarrow -\infty$ the second term in the parentheses can
be neglected and we have the familiar mode functions of the Minkowski space
time. The wavelength of a given mode becomes equal to the horizon size (or
``crosses'' the horizon) when $k\eta = 1$. Inflation proceeds with $\eta
\rightarrow 0$, so the modes with progressively larger $k$ cross the horizon. 
After horizon crossing, when $k\eta \ll 1$, the asymptotics of mode functions
are
\begin{equation}
u_k = \pm \frac{i}{\sqrt{2}k^{3/2}\eta}, {\rm ~~~~or~~~} \varphi_k 
= \frac{u_k}{a(\eta)} = \mp \frac{iH}{\sqrt{2}k^{3/2}} \; .
\label{ModFunc-massles-asimptotic}
\end{equation}
The field variance is given by
\begin{equation}
\langle 0| \phi^2(x) |0\rangle =
\int \frac{d^3k}{(2\pi)^3}\; |\varphi_k|^2 \; .
\label{variance-def}
\end{equation}
and we find in the asymptotic (the careful reader will recognize that this is
already regularized expression with zero-point fluctuations being subtracted)
\begin{equation}
\langle \varphi^2 \rangle = \frac{H^2}{(2\pi)^2} \int \frac{dk}{k} \; .
\label{variance-varphi}
\end{equation}
Defining the power spectrum of the field fluctuations as a power per decade,
$\langle \varphi^2 \rangle  \equiv  \int  P_\varphi(k)\; {d\ln k}$, we find
\begin{equation}
P_\varphi (k) = \frac{H^2}{(2\pi)^2} \; .
\label{PowerSpectrum-varphi}
\end{equation} 

\subsection{Curvature perturbations}

According to Eq.~(\ref{Spatial-curvature}), the three-dimensional curvature of
space sections of constant time is inversely proportional to the scale factor
squared, $^{(3)} R \propto {a^{-2}}$. Therefore, the perturbation of spatial
curvature is proportional to ${\delta a}/{a}$, and this ratio can be evaluated
as
\begin{equation}
\zeta  \equiv \frac{\delta a}{a} = H\delta t = 
H\, \frac{\delta \varphi}{\dot{\varphi}} \; .
\label{zeta-def}
\end{equation}
This allows to relate the power spectrum of curvature perturbations
to the power spectrum of field fluctuations 
\begin{equation}
P_\zeta (k) = \frac{H^{2}}{\dot{\varphi}^{2}} \;\, P_\varphi (k)\; ,
\label{Pphi-Pcurvature}
\end{equation} 
and we find for the power spectrum of curvature perturbations
\begin{equation}
P_\zeta (k) = \frac{1}{4\pi^{2}}\; \frac{H^{4}}{\dot{\varphi}^{2}}\; .
\label{PowerSpectrum-curvature}
\end{equation} 
This very important relation describes inflationary creation of primordial
perturbations, and can be confronted with observations.  The usefulness of
curvature perturbations for this procedure can be appreciated in the following
way:

1. Consider the perturbed metric, Eq.~(\ref{metric-perturbed}). The product
$a(1-\Phi)$ for the long-wavelength perturbations can be viewed as a perturbed
scale factor, i.e. ${\delta a}/{a} = - \Phi$. Comparing this relation with
Eq.~(\ref{zeta-def}) and Eq.~(\ref{theta-phi-md}), we find for the temperature
fluctuations which are of the superhorizon size at the surface of last
scattering
\begin{equation}
\frac{\delta T}{T} = \frac{2}{3}\; \zeta_k \; .
\label{power-T-zeta}
\end{equation}

2. On superhorizon scales the curvature perturbations do not
evolve usually. This fact allows to relate directly the observed
power spectrum of temperature fluctuations to the power spectrum of curvature
fluctuations generated during inflation.

\subsection{Tensor perturbations}

Mode functions of gravity waves (after rescaling by $M_{\rm Pl}/\sqrt{32\pi}$)
obey the same equation as mode functions of massless minimally coupled scalar
\cite{Grishchuk:1975ny}.  Using the result Eq.~(\ref{PowerSpectrum-varphi}) we
immediately find \cite{Rubakov:1982df}
\begin{equation}
P_{T}(k) = 2\; \frac{32 \pi}{M_{\rm   Pl}^{2}} \; P_\varphi(k) =  
\frac{16}{\pi}\; \frac{H^2}{M_{\rm   Pl}^{2}} \; ,
\label{tensor-perturbations}
\end{equation}
where the factor of 2 accounts for two graviton polarizations.

\subsection{Slow-roll approximation}

During inflation, the field $\varphi$ rolls down the potential hill very
slowly.  A reasonable approximation to the dynamics is obtained by neglecting
$\ddot{\varphi}$ in the field equation $\ddot{\varphi}+3H\dot{\varphi} +V'
=0$. This procedure is called the slow-roll approximation
\begin{equation}
\dot{\varphi} \approx -\frac{V'}{3H} \; .
\label{Slow-roll-equation}
\end{equation} 
Field derivatives can also be neglected in the energy density of the inflaton
field, ~$\rho \approx V$
\begin{equation}
H^{2} = \frac{8\pi }{3M_{\rm Pl}^2} \, V \; .
\label{Slow-roll-Hubble}
\end{equation} 
This gives for curvature perturbations
\begin{equation}
\zeta_k \equiv  P_\zeta (k)^{1/2} =
\frac{H^2}{2\pi\;\dot{\varphi}} =  \frac{4H}{M_{\rm Pl}^2}\;
\frac{V}{V'}\; .
\label{Slow-roll-curvature}
\end{equation}

\subsection{Normalizing to CMBR}

As an example, let us consider the simplest model~ $V = \half m^{2}
\varphi^{2}$.~~ We have
\begin{equation}
\frac{V}{V'} =  \frac{\varphi}{2}, ~~~~~~{\rm and}~~~ {H} =
\sqrt{\frac{4\pi}{3}}\frac{m\varphi}{M_{\rm Pl}}  \; .
\end{equation} 
This gives for the curvature fluctuations
\begin{equation}
\zeta_k = \sqrt{\frac{16\pi}{3}}\; \frac{m\, \varphi^{2}}{M_{\rm Pl}^3} \; .
\end{equation}
Using the relation between curvature and temperature fluctuations,
Eq.~(\ref{power-T-zeta}), and normalizing ${\delta T}/{T}$ to the 
measured value at largest ~$l$,~ which is ${\delta T}/{T}
\sim 10^{-5}$ (see Fig.~\ref{fig:CMB}, right panel) we find the restriction on
the value of the inflaton mass in this model:
\begin{equation}
m \approx \frac{\delta T}{T}\; \frac{M_{\rm Pl}}{30} \approx 10^{13}
\;\; {\rm GeV} \; . 
\end{equation}
Here I have used the fact that in this model the observable scales cross the
horizon when $\varphi \approx M_{\rm Pl}$.

\subsection{Slow-roll parameters}

The number of e-foldings ($a = \e^{Ht} \equiv \e^N$) of inflationary expansion
from the time when $\varphi = \varphi_i$ to the end can be found as
\begin{equation}
N(\varphi_i ) = \int_{t_i}^{t_f} H(t) dt = \int \frac{H}
{\dot{\varphi}}\, d\varphi = \frac{8\pi}{M_{\rm Pl}^2} 
\int_{\varphi_e}^{\varphi_i} \frac{V}{V'}\, d\varphi \; .
\label{e-foldings-rsult}
\end{equation}
In particular, in the model Eq.~(\ref{chaotic}) we find that the largest
observable scale had crossed the horizon ($N \sim 65$) when $\varphi_i \approx
3.5M_{\rm Pl}$. All cosmological scales which fit within the observable
universe encompass a small $\Delta \phi$ interval within $M_{\rm Pl} < \varphi
< \varphi_i$.  And inflaton potential should be sufficiently flat over this
range of $\Delta \phi$ for the inflation to proceed.  This means that
observables essentially depend on the first few derivatives of $V$ (in
addition the the potential $V(\phi_0)$ itself). From the first two derivatives
one can construct the following dimensionless combinations
\begin{eqnarray}
&& \epsilon \equiv \frac{M_{\rm Pl}^2}{16\pi}
\left(\frac{V'}{V}\right)^2 \; , \\
&& \eta \equiv \frac{M_{\rm Pl}^2}{8\pi}\frac{V''}{V}  \; ,
\end{eqnarray}
which are often called the slow-roll parameters.

The power spectra of curvature,  Eq.~(\ref{Pphi-Pcurvature}), and of tensor
perturbations, Eq.~(\ref{tensor-perturbations}), in slow-roll parameters can
be rewritten as
\begin{equation}
P_{\zeta}(k) =  \frac{1}{\pi\epsilon }\frac{H^2}{M_{\rm Pl}^2}, ~~~~~~~~~~~ 
P_{T}(k) =  \frac{16}{\pi}\frac{H^2}{M_{\rm Pl}^2}\; .
\label{Power-spectra-slow-roll}
\end{equation}
Comparing these two expressions we find
\begin{equation}
\frac{P_{T}(k)}{P_\zeta(k)} = 16 \epsilon \; .
\label{consistency-relation0}
\end{equation}

\subsection{Primordial spectrum}

In general, the spectra can be approximated as power law functions in $k$:
\begin{eqnarray}
&& P_\zeta(k) =  P_\zeta(k_0) \left(\frac{k}{k_0}\right)^{n_S -1}  \; ,\\
&& P_{T}(k) =  P_{T}(k_0) \left(\frac{k}{k_0}\right)^{n_T} \; .
\label{spectra-powerlow-def}
\end{eqnarray}
To the first approximation, $H$ in Eq.~(\ref{Power-spectra-slow-roll}) is
constant. Therefore, in this approximation, power spectra do not depend on $k$
and $n_S=1$, $n_T=0$. This case is called the Harrison-Zel'dovich spectrum 
\cite{Harrison:1970fb,Zeldovich:1972ij} of primordial perturbations.  
However, in reality, $H$ is changing, and in Eq~(\ref{Power-spectra-slow-roll})
for every $k$ one should take the value of $H$ at the moment when the relevant
mode crosses horizon. In slow roll parameters one then finds (see
e.g. Ref.~\cite{Lidsey:1997np} for the nice overview) 
\begin{equation}
n_S = 1 + 2\eta - 6 \epsilon, ~~~~~~~~~~~  n_T = - 2 \epsilon\; .
\label{n-in-slow-roll}
\end{equation}
We can re-write Eq.~(\ref{consistency-relation}) as a relation between the
slope of tensor perturbations and the ratio of power in tensor to curvature
modes 
\begin{equation}
\frac{P_{T}(k)}{P_\zeta(k)} = -8 n_T  \; .
\label{consistency-relation}
\end{equation}
This is called the {\it consistency relation} to which (simple) inflationary
models should obey.

Different models of inflation have different values of slow-roll parameters
$\eta$ and $\epsilon$, and therefore can be represented in the
($\eta$,$\epsilon$) parameter plane. Using the relations
Eq.~(\ref{n-in-slow-roll}) we see that this plane can be mapped into
$(n_S,n_T)$, or using also Eq.~(\ref{consistency-relation}) into the
$(n_S,r)$   parameter plane, where $r$ is the ratio of power in tensor to
scalar (curvature) perturbations. In this way, different inflationary models
can be linked to observations and constraints can be obtained.

\end{document}